\documentstyle[preprint,eqsecnum,aps,epsf]{revtex}
\newif\iftightenlines\tightenlinesfalse
\tightenlines\tightenlinestrue
\def\eslt{\not\!\!{E_T}}
\def\to{\rightarrow}

\def\tb{\tilde b}
\def\tx{\tilde\chi}
\def\tst{\tilde t}
\def\ttau{\tilde \tau}

\def\tg{\tilde g}

\def\tell{\tilde\ell}
\def\tq{\tilde q}
\def\tw{\widetilde W}
\def\twb{\overline{\widetilde W}}
\def\tz{\widetilde Z}
\def\sgn{\mathop{\rm sgn}}

\begin{document}
%
%
\preprint{\vbox{\baselineskip=14pt%
   \rightline{FSU-HEP-990509}\break 
   \rightline{IFT--P.041/99}\break
   \rightline{BNL-HET-99/12}\break
   \rightline{UH-511-933-99}
}}
\title{TRILEPTON SIGNAL FOR SUPERSYMMETRY
AT THE FERMILAB TEVATRON REVISITED}
\author{Howard Baer$^1$, Manuel Drees$^2$, 
Frank Paige$^3$, Pamela Quintana$^1$ and Xerxes Tata$^4$}
\address{
$^1$Department of Physics,
Florida State University,
Tallahassee, FL 32306, USA
}
\address{
$^2$IFT, Univ. Estadual Paulista, 01405--900 S\~ao Paulo, Brazil
}
\address{
$^3$Brookhaven National Laboratory, 
Upton, NY 11973, USA
}
\address{
$^4$Department of Physics and Astronomy,
University of Hawaii,
Honolulu, HI 96822, USA
}
\date{\today}
\maketitle
\begin{abstract}

Within a wide class of models, the LEP2 lower limit of 95~GeV on the
chargino mass implies gluinos are heavier than $\sim$300~GeV. In this
case electroweak $\tw_1\twb_1$ and $\tw_1\tz_2$ production are the
dominant SUSY processes at the Tevatron, and the extensively examined
isolated trilepton signal from $\tw_1\tz_2$ production assumes an even
greater importance.  We update our previous calculations of the SUSY
reach of luminosity upgrades of the Fermilab Tevatron in this channel
incorporating ({\it i})~decay matrix elements in the computation of the
momenta of leptons from chargino and neutralino decays,
({\it ii})~the trilepton background from $W^*Z^*$ and
$W^*\gamma^*$ production which,
though neglected in previous analyses, turns out to be the dominant
background, and finally,
({\it iii}) modified sets of
cuts designed to reduce these new backgrounds and increase
the range of model parameters for which the signal is observable.
We show our improved projections for the reach for SUSY of both the
Fermilab Main Injector and the proposed TeV33 upgrade. 
We also present opposite sign
same flavor dilepton invariant mass distributions as well as the $p_T$
distributions of leptons in SUSY trilepton events, and comment upon how
the inclusion of decay matrix elements impacts upon the Tevatron
reach, as well as upon the extraction of neutralino masses.

\end{abstract}

\medskip

\pacs{PACS numbers: 14.80.Ly, 13.85.Qk, 11.30.Pb}


\section{Introduction}

The minimal supergravity (mSUGRA) model is a well motivated
framework\cite{sugra} for exploring the experimental consequences of
weak scale supersymmetry. In this model, it is assumed that
supersymmetry breaking occurs in a ``hidden sector'' of the model, and
is then communicated to the observable sector via
gravitational--strength interactions. Motivated by the observed
suppression of
flavor changing neutral currents as well as by the near unification of
gauge coupling constants at $M_{GUT}\simeq 2\times 10^{16}$ GeV, one
assumes a common mass $m_0$ for scalars and a common mass $m_{1/2}$ for
gauginos at scale $Q=M_{GUT}$.  In addition, the soft breaking trilinear
$A$ terms are also unified to $A_0$ at $Q=M_{GUT}$. The soft SUSY
breaking parameters, gauge and Yukawa couplings are then evolved from
$Q=M_{GUT}$ to $Q\sim M_{weak}$ via renormalization group equations. The
magnitude of the superpotential $\mu$ term is determined by requiring
radiative breaking of electroweak symmetry; this latter constraint also
effectively allows one to trade the bilinear $B$ parameter for the
parameter $\tan\beta$. Thus, the entire spectrum of SUSY and Higgs
particle masses (as well as all the couplings) is predicted in terms of
Standard Model (SM) parameters augmented by the SUSY parameter set
\begin{eqnarray}
m_0,\ m_{1/2},\ A_0,\ \tan\beta\ {\rm and}\ \sgn(\mu ).
\end{eqnarray}
This model, along with several others, has been incorporated into the event
generator ISAJET\cite{isajet}.

The negative results of sparticle searches in experiments at the
Tevatron and at LEP have led to significant lower limits on gluino and
chargino masses.  In the mSUGRA framework and in other models with (real
or apparent) gaugino mass unification, the LEP2 limit on chargino mass
$m_{\tw_1}>95$ GeV\cite{lep2} implies that gluinos and most squarks
ought to have masses typically greater than about 300 GeV, so that
strong sparticle production at the Tevatron is expected to be
suppressed.  Then, $\tw_1\twb_1$ and $\tw_1\tz_2$ production is
expected\footnote{This is not to imply that experiments at the
Tevatron ought not to search for gluinos and squarks. Direct search
limits are important since our assumption about the relationships
between gaugino masses could well prove to be incorrect.} to be the
dominant sparticle production mechanism if sparticles are at all
accessible.  It has become increasingly clear that the trilepton signal
from $p\bar{p}\to \tw_1\tz_2 X$ followed by
$\tz_2\to\ell\bar{\ell}\tz_1$ and $\tw_1\to\ell\nu\tz_1$ ($\ell=e,\mu$)
is one of the most promising discovery channels for supersymmetry at
luminosity upgrades of the Fermilab Tevatron
collider\cite{old,bckt,bcpt,bcdpt2,bkl,bk,matchev}.  As an illustration, in
Fig.~\ref{SIGA}, we plot sparticle cross sections as a function of
$m_{\tilde g}$ assuming five generations of degenerate squarks, for {\it
a}) $m_{\tq}=m_{\tg}$ and {\it b}) $m_{\tq}=2m_{\tg}$, assuming
$\mu=+m_{\tg}$, and $\tan\beta =3$ (MSSM parameters with gaugino mass
unification).
The region to the left of the
vertical line is excluded by the LEP2 chargino mass limit $m_{\tw_1}>95$
GeV\cite{lep2}.  It can be seen that electroweak production of
charginos and neutralinos dominates over the
strongly produced $\tg\tg$, $\tg\tq$ and $\tq\tq$ cross sections over
essentially all of parameter space for which $|\mu |\gg M_1,\ M_2$. The
cross section for
$\tw_1\tz_1$ is relatively suppressed because $\tz_1$ is dominantly a
hypercharge gaugino and so couples to $W$ via its suppressed components, 
while the squark or gluino
plus chargino or neutralino associated production reactions (summed over
all sparticle types and shown by dash-dot-dot curves) occur
at smaller rates. These qualitative features hold for both frames
shown.  Similar results are shown in Fig.~\ref{SIGB} for $\mu
=-m_{\tg}$.  In this case, for low values of $m_{\tg}$ with
$m_{\tq}\simeq m_{\tg}$, the strong production cross sections can be
dominant, but only in parameter space regions already excluded by LEP2.
Since opposite sign dilepton or jet plus lepton signals from $\tw_1\twb_1$
production suffer from large Standard Model backgrounds, many
groups have focussed on the clean trilepton signature from $\tw_1\tz_2$
production for which the SM background is expected to be small:
after suitable cuts, this mainly
comes from $p\bar{p}\to WZ^*+X$ or $W\gamma^*+X$, where $W\to \ell \nu$ and 
$Z^*$ or $\gamma^*$ decay leptonically.  

The Tevatron reach in this channel has been extensively examined 
especially for low
values of $\tan\beta$ where the effect of Yukawa interactions is
negligible.
In parts of parameter space of the mSUGRA model (including
those favored by predicting a Big Bang relic abundance of lightest
neutralinos in the cosmologically interesting range\cite{bbrelic}), the
reach is very large because
neutralino leptonic decay rates can be enhanced owing in part to
relatively light sleptons mediating the decay chain.  However, in
other regions of parameter space, these same decays can be suppressed
by large negative interference terms between the $Z$ and slepton
exchange graphs\cite{bt}, and there is no reach even if charginos are
just beyond the LEP bound.

If the mSUGRA parameter $\tan\beta$ is large, then $\tau$ and $b$
Yukawa couplings become non-negligible as well, and as a result,
$\ttau_1$ and $\tb_1$ can be significantly lighter than first and
second generation sleptons and squarks.  Consequently, chargino and
neutralino decays to $\tau$-leptons and $b$-quarks can be enhanced
over decays to their first and second generation counterparts.  If the
decays $\tw_1\to\tau\nu_\tau\tz_1$ and $\tz_2\to\tau\bar{\tau}\tz_1$
are strongly enhanced, trilepton events (where $e$ or $\mu$ are
secondaries from $\tau$ decays) can still occur at a considerable
rate, but then $e$s and $\mu$s will have a much softer energy
distribution making the detection of the signal more
challenging. Hence, unless hadronically decaying taus can be detected
with high efficiency and purity, the Tevatron reach for mSUGRA in
general becomes more limited for high values of
$\tan\beta$~\cite{bcdpt2}.

Because of the importance of this channel, considerable attention was
focussed on optimizing cuts to maximize the Tevatron SUSY discovery
potential during the recent Fermilab Tevatron ``Workshop on SUSY/Higgs
Particles at Run 2.'' Barger, Kao and Li \cite{bkl} pointed out that using 
softer lepton $p_T$ cuts for SUSY trilepton events could improve
the expected SUSY signal to background levels by a significant margin,
particularly for scenarios with large values of $\tan\beta$. In
addition\cite{bk,run2}, the signal level could be increased significantly
relative to background by allowing jetty events into the trilepton
signal sample, and the greatest reach was shown to be obtained via this
{\it inclusive} trilepton channel. 
The Tevatron reach in mSUGRA parameter space was computed in Ref.~\cite{bk}
using soft cuts, and was found to have increased significantly beyond
the results presented in Ref.~\cite{bcpt,bcdpt2}, where harder lepton
cuts (originally devised for the signal at low $\tan\beta$) and a jet
veto were used. 
It has recently been pointed out\cite{matchev} that these studies
neglected contributions from $WZ$ production where the $Z$ boson was allowed
to be off mass shell. These authors included backgrounds 
from $WZ^*$ production using PYTHIA, and looked to optimize cuts for
SUSY trilepton signals over background throughout the mSUGRA model parameter
space. In fact, in much of parameter space, they found hard lepton cuts were
optimal since they could remove more of the $WZ^*$ background that was 
incompletely removed by a $Z$-boson mass veto.

Recently, the event generator ISAJET (v 7.44) has been upgraded to
include decay matrix elements for chargino, neutralino and gluino
three-body decays. In previous versions, while the computation of
branching fractions had included these matrix elements, just phase
space was used in the event generation to determine the energy and
momentum distributions of their decay products.  In practice, because
various kinematic cuts are placed on each of the three signal leptons
in order to extract signal from background, the resultant signal rates
can in fact depend on the momentum and energy distributions of the
decay leptons. This neglect of the matrix element becomes an
especially poor approximation when the mass of the gauge boson mediating
the decay approaches that of the parent neutralino or chargino.
Moreover, as has recently been pointed out by Nojiri and
Yamada\cite{nojiri}, distributions in dilepton invariant mass
$m(\ell^+\ell^- )$ can be sensitive (due to matrix element effects) to
SUSY particle masses and mixings. These can affect not only the
overall $m(\ell^+\ell^- )$ distribution shape, but also the
determination of the distribution endpoint, which yields an important
measure of the mass difference $m_{\tz_2}-m_{\tz_1}$.


In light of these various developments, and the importance of the
trilepton signal to SUSY searches at the Tevatron, we felt it was
worthwhile to update our calculations for the reach of Fermilab Tevatron
upgrades for the mSUGRA model via the trilepton channel. 
We include decay matrix
elements for chargino and neutralino decays in the generation of SUSY
events.
We also perform an exact lowest order calculation of the background 
process $p\bar{p}\to \ell\nu +\ell'\bar{\ell}'X$, including both
$W^*\gamma^*$ and $W^*Z^*$ contributions. The $W^*\gamma^*$ source gives
background rates beyond those considered in Ref.~\cite{matchev}.
We show reach results for both soft and hard lepton cuts,
augmented by some additional cuts designed to eliminate much of
this new background. 

In Section II, we describe our background calculations, and 
list the improved soft and hard cuts that we 
use to reduce these backgrounds
relative to the SUSY signals.
In Section III, we describe in some detail our inclusion of 
decay matrix elements in ISAJET, and examine some distributions which 
reflect their inclusion. 
In Section IV, we perform five case studies of the trilepton signal for
each set of selection cuts introduced in Sec.~II and obtain the best one
for the extraction of the signal. We also study the effect of the matrix
element on the total signal as well as on various distributions.
In Section~V we
present the results of our updated calculations of the
SUSY trilepton reach in the 
$m_0\ vs.\ m_{1/2}$ plane. We conclude in Section~VI with 
a summary of our results and some general remarks.

\section{Backgrounds and selection cuts}

We use ISAJET 7.44 to generate events in the mSUGRA model parameter
space and to generate most of the SM backgrounds.  We use the toy
detector simulation package ISAPLT, assuming calorimetry between
$-4<|\eta |<4$, with an array of calorimeter cells of size
$\Delta\eta\times\Delta\phi =0.1\times 0.262$. We take the
electromagnetic energy resolution to be $0.15/\sqrt{E}$ and the
hadronic calorimeter resolution to be $0.7/\sqrt{E}$ ($E$ in
GeV). Calorimeter cells are coalesced in towers of $\Delta R =0.7$
using the jet finding algorithm GETJET.  Hadronic clusters with
$E_T(j)>15$ GeV are called jets.  Leptons ($e$s or $\mu$s) with $p_T$
of 5~GeV or more are considered to be isolated if the hadronic $E_T$
in a cone about the lepton of $\Delta R=0.4$ is less than 2 GeV.

We have examined the signals and backgrounds using five sets of
acceptance cuts.  A relatively hard set of cuts chosen for the study of
the clean trilepton reach for low values of $\tan\beta$ is taken from
Ref.~\cite{bckt} (note, however, that the lepton isolation criterion
that we use here differs slightly from Ref.~\cite{bckt}). These cuts are
listed in column 2 of Table I, and labelled HC1.  The CDF\cite{cdf} and
D0\cite{d0} collaborations and Mrenna {\it et al.}\cite{mrenna} have
used relatively softer cuts in their analyses. These soft cuts were
advocated in Refs.~\cite{bkl,bk,run2} as being more effective in
eliciting signal from background, especially for large $\tan\beta$,
where many of the signal leptons arise as secondaries from $\tau$ decay,
and are quite soft.  These cuts are listed in column 3 of Table I, and
labelled SC1.  Note that unlike for HC1, a jet veto is not imposed, so
that the signal will be {\it inclusive}, containing both clean and jetty
trilepton events. We list in column 4 the augmented soft cuts that allow
significant suppression of the $W^*\gamma^*$ and $W^*Z^*$ backgrounds;
these are labelled SC2. In column 5, we modify the SC2 cuts imposing
hard lepton $p_T$ requirements, and denote this new set by HC2. For
large values of $\tan\beta$ where $\tz_2 \to \tau\bar{\tau}\tz_1$
dominates the decay of $\tz_2$, trilepton events arise when both $\tau$s
decay leptonically, with a third lepton coming from chargino decay. In
this case, the opposite sign (OS) dilepton pair in the trilepton event
does not have the same flavour (SF) in a quarter of the signal events,
regardless of the origin of the third lepton. Since these secondary
leptons are soft, we use cuts SC1 but veto events with OS/SF pairs to
reject the $Z^*$ and $\gamma^*$ backgrounds; this cut set (which is
introduced to pick up the large $\tan\beta$ signal) is labelled by SC3.

The dominant SM backgrounds are listed in Table II for the five sets of
cuts.  It has recently been pointed out\cite{matchev} that $WZ$
production, where $W\to\ell\nu$ and an off-shell $Z^*\to\ell\bar{\ell}$
is a major background to the SUSY trilepton signal.  Similarly,
$W\gamma^*$ can also lead to significant rates for trilepton
backgrounds.\footnote{The importance of this contribution to the 
background was first
pointed out in Ref.~\cite{mrenna}, which also contains an estimate of its
size. Our results differ significantly from this estimate, especially for
small di-lepton invariant masses, where our calculation gives much higher
rates, due to the $\gamma$-pole.} 
We have used the program MADGRAPH/HELAS\cite{madgraph} to
compute the complete lowest order squared matrix element for the process
$q\bar{q}'\to e^+\nu_e\mu^+\mu^-$~\cite{EC}. The ten contributing Feynman
diagrams
are shown in Fig.~3, and include contributions from
$W^*Z^*$ and $W^*\gamma^*$ production, plus some other
contributions. (Throughout this paper, a star on a particle
indicates that it could be either real or virtual.)
Similarly, we have computed the $q\bar{q}'\to e^+\nu_e e^+ e^-$
background, which includes twenty diagrams.
We have constructed parton-level Monte Carlo programs to
then estimate these backgrounds.  We use ISAJET to calculate the
$t\bar{t}$ and $WZ$ ($Z\to\tau\bar{\tau}$) backgrounds. In addition, we
use ISAJET to calculate backgrounds from $ZZ$ production. For this
latter calculation, we smear each $Z$ decay to $e\bar{e}$ or
$\mu\bar{\mu}$ by a Breit-Wigner distribution to simulate the effect of
the off-shell $Z$ contribution. This cross section is much larger with
HC2 cuts than HC1 cuts because events from $Z^* \to \ell\ell$
while $Z \to \tau\tau$ (where just one of the two taus decay leptonically)
would be mostly removed by the jet veto involved in HC1, 
but would likely pass the HC2 inclusive cuts.

>From Table II, it can be seen that using HC1 or SC1 cuts, there is a
very large background contribution from the $\ell\nu\ell'\bar{\ell}'$
source (denoted by $W^*Z^*$ or $W^*\gamma^*$). 
We plot in Fig.~\ref{MLLWZ}{\it 
a} the dilepton mass distribution for same flavor/opposite sign
dilepton pairs after imposing the SC1 cuts, except for the $Z$
veto. In performing our parton level Monte Carlo for
$\ell\nu\ell'\bar{\ell}'$ production, we cut off $\ell'\bar{\ell}'$
masses below 1 GeV to avoid the singularity from the $\gamma$
propagator. It is also crucial to integrate over the entire range of
invariant mass values for the virtual $W$ contribution since we find
a large background contribution coming from $W^*$'s with invariant mass in the
range of $30-60$ GeV.

There is a large contribution~\cite{EC} to the $3\ell$ background from the $Z$
resonance, much of which is effectively removed by the $Z$ mass veto
cut. 
Here, we expand the $Z$ mass veto cut to include all events with
$m(\ell\bar{\ell})>81$~GeV, which vetos far off-shell $\gamma^*$ or $Z^*$ 
decays.
An even larger contribution comes from the phase space region where
the photon gets close to its mass shell: a ``$\gamma^*$ veto'', must be
imposed to reduce the large $W^*\gamma^*$ contribution at lower
dilepton mass values. This cut should also remove background from
charmonium and bottomonium decays and from $b$ decays\cite{matchev}.  In
Fig.~\ref{MLLWZ}{\it b}, we show the distribution in $m_T(\ell ,\nu )$
after using SC1 plus a $\gamma^*$ mass veto of either 12 or 20 GeV.
Imposing the $m(\ell^+\ell^- )>12$ GeV cut leaves a characteristic
$W$-boson transverse mass distribution, but with a distinctive bulge
around $m_T(\ell ,\nu )\sim 40-60$ GeV due to a large rate for far off
mass shell $W^*$ production. The low mass $W^*$'s are produced in
association with very low mass $\gamma^*$'s. Imposing instead
$m(\ell^+\ell^- )>20$ GeV removes much of the off-shell $W$ bulge and
leaves a more typical distribution in transverse mass.  Much of the
remaining background can be eliminated at some cost to signal by vetoing
events with $60< m_T(\ell ,\nu )<85$ GeV. Insertion of this complete
background process into an event generator with QCD radiation and
detector simulation will broaden the $m_T$ peak somewhat, increasing the
background rate; however, some of this background rate may be decreased
as well due for instance to non-isolated leptons. The cuts SC1 augmented
by the 20 GeV ``$\gamma^*$'' together with the expanded $Z$ veto
and the  $W$ transverse mass veto form our set of cuts
SC2.  We also examine a set of cuts HC2 which is SC1, but with increased
$p_T$ requirements on each of the three leptons together with the $Z$,
$\gamma^*$ and transverse mass vetos.  Corresponding distributions in
$m(\ell^+\ell^- )$ and $m_T(\ell ,\nu )$ are shown in
Fig.~\ref{MLLWZH}. From this Figure, it is clear that for hard lepton
$p_T$ cuts, a ``$\gamma^*$ veto'' $m(\ell^+\ell^- )>12$ GeV is sufficient.
Finally, we examine the cut set SC1, but augmented instead by a veto on
OS/SF dilepton pairs, denoted by SC3. There is, of course, no background
from $W^*Z^*$ and $W^*\gamma^*$ events and, as expected, the dominant
remaining background from $WZ, Z\to \tau\tau$ events drops to about a
fourth.

We also ran $Z+{\rm jets}$ and $W+{\rm jets}$ background jobs.  No
events from 
these two sources passed any of the sets of cuts out of $5\times 10^5$ and
$10^6$ events generated, respectively. These correspond to one event
levels of less than 0.3 and 4 fb, respectively. In runs of $10^8$
$W+{\rm jets}$ events with somewhat different cuts, some $3\ell$ events
could be generated leading to sizeable backgrounds; these sources
always had $b\to c\ell\nu$ followed by $c\to s\ell\nu$, so that these
sources of background could be removed (without any appreciable loss
of signal) by imposing an angular separation cut between the isolated
leptons, giving a background consistent with zero.
Finally, we list in Table II the total background cross section as well as
the minimum signal levels for a $5\sigma$ excess for
integrated luminosity of 2 and 25 fb$^{-1}$ as well as the minimum for a
``$3\sigma$ observation'' with 25 fb$^{-1}$. 
At Run II with 2 fb$^{-1}$ integrated luminosity,
we expect about two (one) events per experiment using cuts SC2 (HC2), for which
the background cross section is $\sim 1$ fb ($\sim 0.5$ fb).
Thus, about 7 signal events will be necessary to establish a 
$5\sigma$ effect at Run II for SC2 cuts. We do not
attempt to quote the increased significance that might be possible by
combining the event samples from the two experiments.
On the other hand, we do not attempt to model experimental
efficiencies, either.

\section{Decay matrix element effects}

The event generator ISAJET 7.44 has recently been 
upgraded to include the effects of
sparticle decay squared matrix elements on the distribution for any
gluino or gaugino to decay to another gaugino plus a fermion-antifermion
pair. Other matrix elements may be incorporated in the future.
Spin correlations are not yet included
in ISAJET, so the SUSY particles are effectively unpolarized, but
these effects are probably less important.
The procedure used is as follows:
\begin{enumerate}
\item We begin by computing a general form for the decay amplitude for
the process $\tilde A\to\tilde B+f+\bar{f}'$, where $\tilde A$ and
$\tilde B$ are gauginos or gluinos, and $f$ and $\bar{f}'$ are SM fermions. 
We construct general amplitudes, {\it i.e.} amplitudes with arbitrary
couplings consistent with the most general Lorentz structure, for decays
via intermediate vector bosons, sfermions, anti-sfermions, scalar and
pseudoscalar particles.  The squared amplitudes including interference
terms are all pre-programmed functions in ISAJET.
\item When each decay branching fraction is calculated in ISAJET, 
the masses and types of each exchanged particle are saved along with the
in general complex coefficients needed to specify the vertices.
In our decay calculations, we include all third generation mixing effects,
Yukawa couplings and tree level decay Feynman graphs\cite{bcdpt2}. 
\item When any type of three body decay is generated, hit-or-miss
Monte Carlo is used to generate an appropriate kinematical set of
decay product four-vectors. 
The maximum of the decay integrand, which is needed for this calculation, is 
calculated the first time via Monte Carlo integration. This result is saved
and updated as better maxima are found.
\end{enumerate}
In addition, ISAJET includes a calculation of partial
widths of particle decays into left or right handed taus independently. 
Taus are then decayed appropriately according to their respective squared
matrix elements.  Spin correlations are neglected. For the case of
sparticles decaying into two taus (such as
$\tz_2\to\tz_1\tau\bar{\tau}$), an average $\tau$ polarization is
used. Thus the effect of tau polarization (which plays an important role
for signals involving hadronically decaying taus \cite{hag,bcdpt2,matchev})
is at least approximately included. QCD corrections to both the signal
and the background are neglected in our analysis.

The efficiency with which leptons from $\tw_1$ and $\tz_2$ decays pass
our cuts depends directly on their transverse momentum spectra.  These
are shown for three mSUGRA cases in Figs.~\ref{PTL1}, \ref{PTL2} and
\ref{PTL3}, where we compare phase space distributions (dashed
histograms) with predictions with the matrix element included (solid
histograms) after the soft cuts SC1. Since our purpose is to illustrate
the effect of the matrix elements for $\tw_1$ and $\tz_2$ decays, these
distributions are shown for just $\tw_1\tz_2$ production, and not for
the case where all SUSY processes are included. In Fig.~\ref{PTL1} we
illustrate these for case A with ($m_0,\ m_{1/2},\ A_0,\ \tan\beta ,\
\sgn(\mu ))=(100,\ 170,\ 0,\ 3,\ 1)$ (where dimensionful quantities are
in GeV units). In this case, $m_{\tz_2}=117$ GeV, $m_{\tz_1}=62$ GeV and
$m_{\tell_R}=126$ GeV, so that slepton exchange dominates leptonic
$\tz_2$ decays. Because the $\tell_R$ mass is so close to $m_{\tz_2}$ it
would seem reasonable to expect that the inclusion of the matrix element
would tend to enhance the rate for configurations where the third lepton
is very soft (since this brings the intermediate slepton closest to its
mass shell).  We see, however, that the matrix elements predict very
similar $p_T$ spectra for {\it all three} leptons compared to just phase
space, so that in this case the phase space approximation works
surprisingly well. A closer look at the squared matrix element for the
``slepton-mediated $\tz_2$ decay'' reveals that it actually {\it
vanishes} in the limit that the momentum of either of its daughter
leptons goes to zero, completely nullifying the enhancement expected
from the propagator.  For case B in Fig.~\ref{PTL2}, with
($m_0,\ m_{1/2},\ A_0,\ \tan\beta ,\ \sgn(\mu ))=(250,\ 175,\ 0,\ 3,\
1)$, there is a large negative interference between $Z$ and
slepton mediated decay graphs. In this case, for the two highest $p_T$
decay leptons, the results with exact decay matrix elements
give a slightly softer $p_T$ spectrum, while that of the third lepton is
somewhat harder.  Fig.~\ref{PTL3} shows results for case C where we
choose ($m_0,\ m_{1/2},\ A_0,\ \tan\beta ,\ \sgn(\mu ))=(500,\ 200,\ 0,\
3,\ -1)$ for which $m_{\tz_2}=173$ GeV, $m_{\tz_1}=86$ GeV and
$m_{\tell_R}=507$ GeV. The parameters are chosen such that $\tz_2$ can
decay through a nearly on--shell $Z$ boson; the leptons from $\tz_2$
decays are then usually quite energetic. As a result, the inclusion of
the matrix element results in a considerably harder $p_T$ distribution
for the three leptons relative to expectation based on phase space
alone.

In Fig.~\ref{POINT1}, we show the mass distribution of opposite sign
(OS) same flavor dileptons produced in $\tw_1\tz_2$ events for the same
mSUGRA point as in Fig.~\ref{PTL1} (case A), where virtual slepton exchange
dominates the $\tz_2\to\tz_1\ell\bar{\ell}$ decay. Again, the phase
space distribution is denoted by dashes, while the exact results are
solid.  We use the SC1 cuts 
described in Section II in this figure (except for removing the
$m(\ell\bar{\ell})$ cut around $M_Z$), but include no background, and
normalize to unity. For this point, the invariant mass of $\ell^+
\ell^-$ pairs from $\tz_2$ decays is bounded by $m_{\tz_2}-m_{\tz_1}=
55.3$ GeV; the few events at larger invariant masses arise from the
two-fold ambiguity present when all three leptons are the same flavor.
For this case, there is hardly any shift in invariant mass between the
two cases.

In Fig.~\ref{POINT2}, we show the $m(\ell\bar{\ell})$ distribution for 
case B also shown in Fig.~\ref{PTL2} where the intermediate sleptons are quite
heavy. In this case, there is a distinct shift of the distribution
towards lower invariant masses when the decay matrix element is
included. This arises from a cancellation between $Z$ and slepton
mediated decay graphs which actually suppress the invariant mass
distribution near its kinematic limit, as first noted by Nojiri and
Yamada\cite{nojiri}. This situation can lead to potentially greater
uncertainties in measuring the $\tz_2$-$\tz_1$ mass difference.

Finally, in Fig.~\ref{POINT3}, we show the $m(\ell\bar{\ell})$ distribution
for case C where sleptons are very heavy, but the intermediate $Z$ boson
in the decay process can be nearly on mass shell.  As might be expected,
the $Z$ pole pulls the
dilepton invariant mass towards $M_Z$, and clearly
illustrates the inadequacy of assuming a pure phase space decay distribution.
In this case, the peak in the signal distribution will merge with 
the background contribution from off-shell $Z$ decays in $WZ$
production, and much of the signal will be eliminated by the $Z$-veto cut.
The phase space approximation will thus result in a significant
overestimate of the signal in this case.

\section{Five Case Studies}

In order to compare the five sets of cuts we have performed five case
studies, each case being characterized by a qualitative feature of
sparticle production mechanism or sparticle decay pattern as described
below.  These scenarios were first examined at the Fermilab Run II
Workshop on SUSY/Higgs physics. The first four are realized within the
mSUGRA framework, while in the fifth one non-universal soft SUSY
breaking Higgs masses were chosen at the GUT scale to realize a ``low
$\mu$ scenario''.  The model parameters for each of these cases is
listed in Table~III along with several sparticle masses and production
cross sections for relevant SUSY production processes.

We note the following features of each case study point.
\begin{itemize}

\item {\bf Case 1:} The mSUGRA parameters for this point lie in the
cosmologically favored region of parameter space\cite{bbrelic}, 
giving rise to a
reasonable relic density of neutralinos. The dominant production mechanisms
at the Tevatron are $\tw_1\twb_1$ and $\tw_1\tz_2$ production.
For this case,
$\tz_2\to \ell\tell_R$ at $\sim 100\%$, so a large rate for clean trilepton
events is expected, and decay matrix element effects are unimportant.

\item {\bf Case 2:} This parameter space point is selected to have a
large value of $\tan\beta =35$ so that $\tw_1\to\ttau_1\nu_\tau$ and
$\tz_2\to\ttau_1\tau$ occur with a branching fraction of $\sim
100\%$. The dominant production cross section is again
$\tw_1\twb_1$ and $\tw_1\tz_2$ production. Here, we anticipate
that an inclusive trilepton signal can be more effectively extracted
with relatively soft lepton $p_T$ cuts, since the detected leptons
typically come from $\tau$ decays. Events containing a mixture of $(3-
n)$~$e$s or $\mu$s, together with {\it n} $\tau$-jets should also exist ($1 \leq
n \leq 3$).
 
\item {\bf Case 3:}
This parameter space point is also chosen with large $\tan\beta$, but the
$A_0$ parameter was chosen so that relatively light $\tst_1$, $\tb_1$
and $\ttau_1$ are generated.  $\tw_1\to\ttau_1\nu_\tau$ and
$\tz_2\to\ttau_1\tau$ occur again at $\sim 100\%$, but the masses of
$\tw_1$ and $\tz_2$ are about 20 GeV smaller than in case 2, so that
the trileptons should occur at about twice the rate as in case
2. Moreover, the rather large $\tst_1\bar{\tst}_1$ production cross section
may yield an observable $\tst_1$ signal but we will not investigate this
here.  Once produced, $\tst_1\to b\tw_1$ with a branching fraction $\sim
100\%$, but since $\tw_1\to\ttau_1\nu_\tau$, hard electrons or muons are not
generated in the $\tst_1$ cascade decay. The cross section for all
flavors of squarks and for gluinos is about 50\% of the total cross section.

\item {\bf Case 4:} This parameter choice leads to
$\tw_1\twb_1$, $\tw_1\tz_2$ and $\tst_1\tst_1$ production as the
main SUSY processes.  It
could yield a sample of high $p_T$ trilepton events. One may also
search for $\tst_1\tst_1$ production where $\tst_1\to b\tw_1$ with
$\tw_1\to \ell\nu_\ell\tz_1$, but this is beyond the scope of our
study. Since charginos and neutralinos decay via three body modes, the
decay matrix element effect may be important.

\item {\bf Case 5:} This point was chosen to have rather large Higgs masses
at the GUT scale, so that scalar universality is broken. The $\mu$ parameter
is relatively small so that the lower lying charginos and neutralinos have
a substantial higgsino component. In this case, $\tw_1\twb_1$,
$\tw_1\tz_2$ and $\tw_1\tz_3$ all occur at large rates.
$\tz_2\to ee\tz_1$ occurs with a 3\% branching ratio, but $\tz_3\to\ttau_1\tau$
at $\sim 100\%$. This case may lead to clean, hard trileptons from
$\tw_1\tz_2$ production, but
also contain a soft trilepton component from $\tw_1\tz_3$ production.
Decay matrix elements can again be important in this case.

\end{itemize}

\subsection{Observability of the SUSY Signal}

The total cross section for production of {\it all} sparticles
is shown in Table~III. We also list here the percentage of
cross section for various relevant sparticle production mechanisms.
We see that while chargino and neutralino production dominate in cases
1, 2 and 5, the production of squarks and gluinos (mainly stops and
sbottoms) is important in cases 3 and 4.

The trilepton event cross sections after cuts for cases 1-5 introduced
above are shown in Table IV for the five sets of cuts introduced in
Table ~I. For the soft cuts the signal with SC1 exceeds SC2 because of
the additional $\gamma^*$ and $W$-veto requirements. While these
reduce the signal by 40-50\%, the corresponding reduction in the
background is by a factor of 35 (mainly due to the reduction of the
$W^*Z^*$ and $W^*\gamma^*$ backgrounds).  Comparing with the BG rates
from Table~II, we see that with 2 fb$^{-1}$ none of the cases are
observable at the $5\sigma$ level using cuts SC1!
Using cut set SC2, only case 1 is visible with 2 fb$^{-1}$ while cases
1, 3 and 4 should be observable at TeV33. 
For the
hard cuts, it is clear that the set HC2 performs better than the set
HC1: at TeV33, cases 1 and 4 are observable using HC2
in contrast to just case 1 via HC1.\footnote{The reader may wonder
that for cases 2 and 3 the signal with HC2 cuts is larger than that
for HC1 cuts where all the requirements (except the jet veto) are
milder. We have checked that in these cases
bulk of the cross section comes from SUSY reactions (which may contain
jets) other than $\tw_1\tz_2$ production.  In contrast, for case 1, 83\%
(71\%) of the signal with HC1 (HC2) cuts originates in the $\tw_1\tz_2$
process.
}
Notice also that with SC3 cuts none of the signal cases are observable
with just 2~fb$^{-1}$ of integrated luminosity, and only case 3 is
observable at TeV33. This is, of course, to be expected for the low
$\tan\beta$ cases 1 and 4, and for case 5 since neutralinos then decay
to OS/SF dileptons.  The point, however, is that although these cuts
reduce the background very significantly, the corresponding loss of
the signal is simply too large to make this strategy very promising
even at TeV33. While these cuts offer the best signal to background
ratio for cases 2 and 3, the significance $S/\sqrt{B}$ is worse than
that for cuts SC2 in case 2, and in case 3 the two sets of cuts give
about the same statistical significance. In some regions of parameter
space cuts SC3 may thus be useful as a diagnostic once a signal has
been found, but they do not increase the reach.

We conclude that over a large part of the parameter space,
the inclusive soft cuts set SC2
appear to provide the best strategy for extracting the trilepton signal,
since the dominant $W^*Z^*$ and $W^*\gamma^*$ backgrounds are
largely reduced by the three extra cuts included here. It may, however, be
that the cuts HC2 may provide a somewhat better reach for sufficiently
large values of $m_{1/2}$ (and sometimes a better signal to background ratio).

Finally, to illustrate the impact of the matrix element, we also list in
Table~IV the cross section for the inclusive signal with soft cuts set
SC1 but {\it without} the decay matrix elements included; {\it i.e.} as
it was computed prior to the release of ISAJET 7.44. Of course, there is
no change for the first three cases as three body decays of the chargino
and neutralino were not important, but for cases 4 and 5, the inclusion
of decay matrix elements decreases the predicted observable event rate
by 10-15\%.  We have also shown the signals for cases A, B and C
corresponding to mSUGRA parameters in Figs.~\ref{PTL1}, \ref{PTL2} and
\ref{PTL3}, respectively. Although the cross section is unobservably
small except in case A, we have shown these as additional examples that
enable the reader to gauge the importance of the matrix element effect
for the evaluation of the reach.  We see that as expected from
Figs.~\ref{PTL1} and \ref{PTL2}, the matrix element has little effect
for cases A and B. For case C, Fig.~\ref{PTL3} would suggest that there
should be a substantial increase in the cross section when matrix
elements are included as these cause a substantial hardening of the
lepton $p_T$ distributions; we see, however, that the cross section is
substantially reduced. This is because, as we saw from
Fig.~\ref{POINT3}, the matrix element causes the masses of the dileptons
from $\tz_2$ decays to peak near to $M_Z$, so that these events are
removed by the $Z$-veto. We conclude that at least for the soft cuts
considered here, matrix element effects generally do not alter the
signal cross section by more than 10-20\% except for small regions of
parameter space where a three-body decay is close to becoming a quasi
two body decay. As pointed out in Ref.\cite{nojiri} this is not,
however, the case for dilepton mass (and presumably other) distributions
which we consider next.

\subsection{Dilepton invariant mass distributions}

It is known that the mass distribution of same flavor, opposite sign
dileptons can provide important information about neutralino, and
possibly also slepton, masses. This distribution has recently been
studied in Ref.\cite{nojiri} where interesting effects arising from
the matrix element have been pointed out. The magnitude and nature of
these effects is sensitive to underlying parameters, and so can (in
some cases) provide an additional tool to obtain these. Here, we
examine this distribution for the five cases that we have examined in
detail, incorporating the inclusive soft cuts SC2 which for the five
case studies yield the largest reach at luminosity upgrades of the
Tevatron. In several of the cases the leptons arise as secondaries
from the decay of the parent tau that is produced via the decay of
$\tz_2$. In this case, the dilepton mass distribution will be squeezed
to lower mass values, and the sharp edge that directly provides
information about sparticle masses is washed out. Nevertheless, we
show these distributions as they illustrate what might be expected in
such SUSY scenarios even though they only roughly yield information
about neutralino or slepton masses.

In Fig.~\ref{MLL}, we show the resulting dilepton mass distribution
after the cuts SC2 and with the backgrounds included. The frames are
labelled by the corresponding case number.  The final histogram labelled
BG shows the distribution from the various backgrounds shown in Table II
with the same set of cuts. The rise at low $m(\ell\bar{\ell})$ is due to
the tail of the $W^*\gamma^*$ background discussed earlier. This will
obscure the determination of the lower end point of this distribution. 
The sharp
cut off at the high end is from the $Z$-veto cut --- without this, the
distribution has a huge peak at  $m(\ell\bar{\ell}) = M_Z$. 

For case 1, the trilepton signal is large and a mass endpoint may be
visible even at Run 2 where $\sim 14$ signal events are expected. In
this case, $\tz_2\to \ell\bar{\ell}\tz_1$ via a real $\tell_R$, so an
edge is expected at
\begin{eqnarray*}
m_{\ell\bar{\ell}}^{max}=m_{\tz_2}\sqrt{1-{m_{\tell_R}^2\over m_{\tz_2}^2}}
\sqrt{1-{m_{\tz_1}^2\over m_{\tell_R}^2}}\simeq 45\ {\rm GeV},
\end{eqnarray*}
and is clearly visible in the plot. With an integrated luminosity of
25~fb$^{-1}$ a moderately precise measurement of the end point should be
possible. 

For case 2, there should be a similar edge -- but in the
$m(\tau\bar{\tau})$ distribution -- at 54.5 GeV. Dileptons from the
subsequent $\tau$ leptonic decays should also respect this bound, but
with a softened mass distribution. Indeed this appears to be the case
when compared against the pure background distribution shown in the last
frame: there is some signal enhancement beyond the SM expectation
for dilepton masses ranging between about 25--50 GeV. We note, however,
that when two of the three leptons are secondaries of taus from the
neutralino in $\tw_1\tz_2$ production, there is no reason to expect that
the leptons from $\tz_2$ decay should have the same flavour. In fact in
events where a tau each decays into $e$ and $\mu$, the same flavour pair
necessarily comes in association with a lepton from elsewhere (here from
chargino decay) and so should not be expected to respect this end point
(except that the mass is small as the daugter lepton from tau will
usually be soft).  Furthermore, for case 2, production of charged
sleptons and/or sneutrinos and, to a lesser extent heavier charginos and
neutralinos, make significant contributions to the trilepton signal,
frequently without any $\tz_2$ in the cascade. We thus expect no sharp
edges and no real structure to the $m(\ell\bar{\ell})$ distribution for
this case.  Moreover, the event sample will be quite limited even with
25 fb$^{-1}$ of integrated luminosity since less than 20 signal events (and a
comparable number of background events) make up the plot.  Extraction of
information on neutralino or slepton masses from this distribution
appears to be very difficult.  It might, however, be interesting to examine
the possibility of constructing the mass edge using identified
hadronically decaying $\tau$'s; since this clearly depends on detector
capabilities, we have not attempted to do so.

Case 3 should be very similar to case 2 as here, $\tz_2$ again
essentially always decays via $\tz_2\to\tau\ttau_1$. In this case we
expect $m(\tau\bar{\tau})$ should be bounded by 47 GeV. As in case 2,
the trilepton signal originates in many different SUSY sources. Aside
from the usual $\tw_1\tz_2$ production, third generation squarks and
slepton/sneutrino production contribute significantly to this sample
with smaller contributions from other SUSY reactions. It is interesting
to see that the dilepton mass reconstruction again (mostly) respects the
bound $m(\ell\bar{\ell}) \leq m_{\tz_2}-m_{\tz_1}$; Cases 2 and 3
clearly show the care that must be exercised before inferring the origin
of SUSY trilepton events {\it even though the $m(\ell\bar{\ell})$
distribution clusters in a limited range.} Extraction of precision mass
information from this distribution again seems very difficult.

In case 4, $\tz_2\to \ell\bar{\ell}\tz_1$ via virtual particles,
so we expect $m(\ell\bar{\ell})$ to be bounded by
$m_{\tz_2}-m_{\tz_1}=54$ GeV. The mass edge, though not as sharp as in
case 1, is evident in the figure. It should, of course, be remembered
that the signal cross section is just 2.1~fb so that a substantial
integrated luminosity will be needed to extract $m_{\tz_2}-m_{\tz_1}$
with any precision.

Finally, in case 5, dileptons can occur from $\tz_2$ via virtual
sparticles or virtual $Z$, and also from $\tz_3$ decays via a real
$\ttau_1$.  In the $m(\ell\bar{\ell})$ distribution shown in the figure
we expect a mass edge at $m_{\tz_2}-m_{\tz_1}=44$ GeV.  The decay
$\tz_3\to\tau\ttau_1$ will likewise have a $m(\tau\bar{\tau})$ edge at
39 GeV with a correspondingly softer dilepton mass distribution. 
While
there is a slight dip in the distribution at $\sim 55$~GeV, a clear mass edge
does not appear evident.  Note that with 25 fb$^{-1}$ of integrated
luminosity, only about 15 signal events will be used to create this plot, so
the statistical sample will be very limited.
This scenario is further complicated by the fact that both
$\tw_1\tz_2$ and $\tw_1\tz_3$ production is substantial. 
Extraction of masses, though possible in principle,
appears to be difficult. It would be interesting to examine whether the
detection of a signal with taus identified via their hadronic decays
would make it possible to identify the additional presence of $\tz_3$ in
the SUSY event sample.

\section{Reach of Tevatron for mSUGRA via the inclusive isolated
trilepton channel}

In this section, we use the augmented  cut sets SC2 and HC2 (which
effectively removed the bulk of the new backgrounds evaluated here)
to calculate the SUSY
trilepton signal, and thereby determine the reach of Tevatron
upgrades in the parameter space of the mSUGRA model.
We present results for $\mu >0$, since much of the parameter space
for $\mu <0$ and low $\tan\beta$ is ruled out by the recent LEP2 bound
on the mass of the
Higgs boson. For large values of $\tan\beta$, the reach plots become
similar regardless of the sign of $\mu$. 
We note that values of $\mu <0$ also seem to be disfavored from
comparing model predictions
for $b\to s\gamma$ decay rates\cite{bb} 
to experimental results from CLEO\cite{cleo} and ALEPH\cite{aleph}.

In earlier studies \cite{bckt,bcpt,bkl}, reach results were presented for the
Fermilab Tevatron for the low $\tan\beta =2$ value. 
All the observable regions for $\tan\beta =2$ are now ruled out
by the LEP2 result that 
$m_h>95$ GeV.\footnote{The LEP2 bound\cite{lep2higgs} is obtained for the
Higgs boson of the Standard Model. It should, however, also be
applicable to $h$ since for low values of $\tan\beta$, $h$ is essentially
the SM Higgs boson.} 
Increasing $\tan\beta$ by just one unit, to $\tan\beta =3$, typically
raises the value of $m_h$ by 10 GeV, placing it beyond the current reach
of LEP2 Higgs searches. 

To illustrate the reach of the Tevatron for low
$\tan\beta$ values not accessible at LEP2, we evaluate the
observability of the SUSY trilepton signal for $\tan\beta=3$ and show
our results in Fig.~\ref{PAM3N} for  
$\tan\beta =3$ in the $m_0\ vs.\ m_{1/2}$ plane, for $A_0=0$ and
$m_t=175$ GeV.  The black regions are excluded by theoretical
constraints: either electroweak symmetry is not correctly broken, or the
lightest SUSY particle is charged or colored. The grey regions are
excluded by constraints from LEP2 that $m_{\tw_1}>95$ GeV\cite{lep2}.
We have scanned the points on a
(25~GeV~$\times$~25~GeV) grid on this plane to see whether the SUSY
signal is observable above background at the $\geq 5\sigma$ level.
At parameter space points marked with a black square the signal cross
section after cuts exceeds 3.62~fb, and so should be observable at
Tevatron Run II with
SC2 cuts, assuming an integrated luminosity of 2 fb$^{-1}$.
At points marked by an
open square the cross section exceeds 1.02 fb, and hence are
considered to give a detectable signal at TeV33 at $5\sigma$ level, 
assuming 25 fb$^{-1}$ of integrated luminosity. Also shown as 
diamonds are points accessible at the $3\sigma$ level at TeV33. While
this is not a discovery limit, it gives the reader an idea of the
parameter range where
tantalizing hints of SUSY might be possible.
Points where the signal is undetectable even at this level
with 25 fb$^{-1}$ of integrated luminosity are denoted by dots. 

In this case, the Tevatron reach with just 2 fb$^{-1}$ extends to
$m_{1/2}= 225$ GeV.  The reach of the Tevatron with 25 fb$^{-1}$
typically extends to $m_{1/2}=250$ GeV, corresponding to $m_{\tg} \sim
660$ GeV and $m_{\tw_1}\sim 200$ GeV.  Coincidentally, this is the same
as the $\tan\beta=2$ reach\cite{bcpt,bcdpt2} quoted for the {\it clean} $3\ell$
channel (but neglecting the $W^*Z^*$ and $W^*\gamma^*$ backgrounds). 
For $\tan\beta =3$, the reach in
$m_0$ is rather limited, and cuts off sharply at $m_0=200-225$ GeV,
where the $\tz_2$ leptonic branching fraction severely decreases due to
interference effects.  Some reach for TeV33 is recovered at very large
$m_0> 500$ GeV for small $m_{1/2}$ where the $Z$ pole begins to dominate
the $\tz_2$ leptonic branching fraction.

In Fig.~\ref{PAM35N}, we show the corresponding reach of the Tevatron
using SC2 for a large value of $\tan\beta =35$. In this case, for low
values of $m_0$, $\tz_2$ and $\tw_1$ decay dominantly to $\tau$ leptons
due to the dominance of decays mediated by (real or virtual)
$\ttau$'s. None of the parameter space points examined
yields an observable signal at the Tevatron with just 2 fb$^{-1}$.
A reach significantly beyond the current LEP2 limits is possible if 25
fb$^{-1}$ of data is accumulated, especially at large $m_0$.  In
addition, many of these same points plus others are accessible via the
$\eslt + {\rm jets}$ and $\ell\ell\tau$ or $\ell\tau\tau$ signals
discussed in Ref.~\cite{bcdpt2}.

Similar reach plots\footnote{We have explicitly checked that even for
the $\tan\beta=35$ case at TeV33 there is very little reach via SC3
cuts.} are shown for the same parameter planes using the cuts HC2 in
Figs.~\ref{PAM3NH} and \ref{PAM35H}.  Qualitatively, much of the reach
is similar, although HC2 work better in Fig.~\ref{PAM3NH} around
$m_0\sim 200$ GeV and large $m_{1/2}$ (where the signal leptons are
expected to be hard), whereas SC2 work better at very low $m_0$ (where
the mass difference between the sleptons and $\tz_1$ is small). In
Fig.~\ref{PAM35H}, we see that SC2 work better at low $m_0$ where the
trileptons occur dominantly from cascade decays involving $\tau$
leptons. For large $m_0$, the $\tz_2$ decays are dominated by the $Z$
exchange graph. In this case, HC2 work slightly better at 
very large $m_{1/2}$. 
Further optimization of cuts is possible as already noted.

\section{Summary and Concluding Remarks}

Within the mSUGRA model (or any framework with a common gaugino mass at
a high scale $\sim M_{GUT}$), the LEP2 bound $m_{\tw_1} \geq 95$~GeV
translates to $m_{\tg} \agt 300$~GeV. In this case, electroweak
production of charginos and neutralinos is the dominant sparticle
production process at the Tevatron, and the trilepton signal from
$\tw_1\tz_2$ production assumes even greater importance than in the
past.  It has recently been emphasized that the SUSY reach via this
channel can be increased by softening the $p_T$ requirements on the
leptons, and further, by {\it not} imposing a jet veto on these
events. In contrast to these developments that enhance the reach
relative to previous projections, it has also been pointed out that
$WZ^* \to 3\ell$ production, which had been omitted in previous
analyses, is a major source of SM background to the trilepton
signal; the inclusion of this new background of course reduces the
reach. Moreover, as the $m(\ell\bar{\ell})$ spectrum from SUSY events
[see Figs.~\ref{POINT1}-\ref{POINT3}] extends to very low values, SM
background from $W\gamma^*$ background (which had never been evaluated
before) needed to be incorporated.
Finally, previous studies of the trilepton signal ignored the
effect of the chargino and neutralino decay matrix elements in the
evaluation of the energy and angular distributions of the leptons. While
these do not alter total rates, the rates after experimental cuts can be
changed. Recently, the tree level decay matrix elements have been
included  in the event generator ISAJET which is often used to compute
the SUSY signal. In view of these developments and also the importance
of the SUSY trilepton signal, we felt a re-assessment of the signal was
warranted. An improved calculation of the Tevatron reach via this
channel is the main subject of this paper.

For the soft inclusive cuts used here, incorporation of the matrix
elements, for the most part, changes the signal cross sections by less
than $\pm 15$\% so that the conclusions about signal levels from other
studies where these are ignored should be qualitatively correct.  This
is not to say matrix element effects are always unimportant. First,
there are regions of parameter space where the matrix elements do
qualitatively change the distributions ({\it e.g.} case C in Sec.~III),
and hence, signal cross sections. More importantly, matrix element
effects can considerably distort the distribution of same flavor
opposite sign dileptons in SUSY trilepton events from which important
information about neutralino, and perhaps also slepton, masses may be
obtained. This issue may be even more important for SUSY studies at
$e^+e^-$ linear colliders or at the LHC where, because of the clean
environment and/or the large rates, the extraction of sparticle masses
is an integral part of the SUSY program.

Turning to our main purpose, we have shown that augmenting previously
proposed soft or hard
cuts with additional cuts designed to reduce $W\gamma^*$ and $WZ^*$
backgrounds can still lead to substantial regions of parameter space
where the trilepton signal should be observable at Tevatron upgrades.
Our updated projection of the Tevatron
reach via the trilepton channel is summarized in Figs.~\ref{PAM3N},
\ref{PAM35N}, \ref{PAM3NH} and \ref{PAM35H} 
for both low and high $\tan\beta$,
as well as hard and soft lepton $p_T$ cuts. Soft cuts perform slightly better
for the anticipated luminosity of Run II. However, with 25~fb$^{-1}$
of data, the hard cuts can extend the reach towards larger values of
the gaugino mass parameter $m_{1/2}$, for some values of the scalar
mass parameter $m_0$. It is noteworthy that a $5 \sigma$ signal at Run
II requires a signal to background ratio of at least 3.5 to 1 (5 to 1 
for the hard cuts). In contrast, at TeV33 and with soft cuts, signal 
and background are about equal at the $5 \sigma$ discovery limit:
a good knowledge of the background normalization is then 
very important.

Overall we find that, while the reach is limited in that there are
large ranges of parameters where experiments at the Tevatron may see
no signal in this channel even if $m_{\tw_1}$ is just beyond the LEP2
bound, experiments at Run~II should be able to probe significant
parameter ranges not accessible at LEP if $\tan\beta$ is small.  For
$\tan\beta=35$ (or larger) there is very little signal in this channel
at Run~II, although some ranges of parameters become accessible with
25~fb$^{-1}$ of integrated luminosity.  This underscores the
importance of other channels for SUSY detection, particularly when
$\tan\beta$ is large. The standard $\eslt$, the $\eslt$ plus tagged
$b$, the $\ell\ell\tau$ and $\ell\tau\tau$ (where $\tau$ is identified
via its hadronic decays) channels are especially important
\cite{bcdpt2,matchev} in this regard. Since we do not know what nature
has in store for us, we reiterate the need to develop techniques to
efficiently tag $\tau$ leptons via their hadronic decays to ensure
that new physics signals (not necessarily from supersymmetry) do not
evade detection at luminosity upgrades of the Tevatron. The
development of tau triggers does not appear to be as essential at this
point.

Before closing, we mention that it may be possible to further optimize
the cuts once a SUSY signal has been detected. For instance, if data
indicate that the upper end point of the dilepton mass distribution happens to
be well below $M_Z$, it may be possible to significantly reduce the
dominant $WZ^*$ background (thereby increasing the significance of the
signal) by widening the $Z$-veto window. 
The point, of course, is that once we have a qualitative picture of the
signal, further optimization is likely to be possible.

{\it Note added:} After completion of this work, a revised version of
Ref. \cite{bk} appeared which also presented a calculation of the
$W^*Z^*$ and $W^*\gamma^*$ backgrounds. Our background calculations
agree with those of Ref. \cite{bk} when the entire phase space is
integrated over.  Updated work by Matchev and Pierce \cite{mp} also
includes these backgrounds.  Ref. \cite{bk} also includes (sub-dominant)
background contributions from off-shell vector boson decays to $\tau$s
that we have neglected in our analysis.  The conclusions for the reach
obtained in both these studies appear to be in good qualitative
agreement with ours.

%
\acknowledgments We thank C. Kao and V. Barger for discussions and
collaboration on the Fermilab Run II SUGRA subgroup report, which led to
this project.  We also thank P. Mercadante and Y. Wang for numerical
checks on some of our results, and K. Matchev and D. Pierce for
discussions concerning the $WZ^*$ background.  We are grateful to C.~Kao and
K. Matchev for comparisons that made us realize the importance of the
contributions to the trilepton background from $WZ^*$ and $W\gamma^*$
production with the $W$ boson far off-shell: contributions from $M_{W^*} <
70$~GeV, which had been ignored in an earlier version of this manuscript,
significantly increase the background for low values of $m_{\ell\ell}$ and
led us to increase the dilepton mass cut for the SC2 case to $m_{\ell\ell}
\geq 20$~GeV.  The work of M.D. was supported by FAPESP (Brazil).  This
work was supported in part by the U. S. Department of Energy under
contract number DE-FG02-97ER41022, DE-AC02-98CH10886, and
DE-FG-03-94ER40833. 

\newpage

\begin{table}
\begin{center}
\caption{Hard (HC1, HC2) and soft (SC1, SC2, SC3) cuts for Tevatron
SUSY trilepton searches. See Section II for discussion.}
\bigskip
\begin{tabular}{lccccc}
cut & HC1 & SC1 & SC2 & HC2 & SC3 \\
\hline
$p_T(\ell_1)$ & $>20$~GeV & $>$11~GeV & $>$11~GeV & $>20$~GeV & $>$11~GeV \\
$p_T(\ell_2)$ & $>$15~GeV & $>$7~GeV & $>$7~GeV & $>$15~GeV & $>$7~GeV \\
$p_T(\ell_3)$ & $>$10~GeV & $>$5~GeV & $>$5~GeV & $>$10~GeV & $>$5~GeV \\
$|\eta (\ell_{1,2/3})|$ & $<$2.5 & $<$1.0,2.0 & $<$1.0,2.0 & 
$<$1.0,2.0 & $<$1.0,2.0 \\
$ISO_{\Delta R=0.4}$ & $<$2 GeV & $<$2 GeV & $<$2 GeV & $<$2 GeV & $<$2 GeV \\
$\eslt$ & $>$25 GeV & $>$25 GeV & $>$25 GeV & $>$25 GeV & $>$25 GeV \\
$Z-veto$ & 83-99~GeV & 81-101~GeV & $<$81~GeV & $<$81~GeV 
& --- \\
$N(j)$ & 0 & --- & --- & --- & --- \\  
$m(\ell\bar{\ell})$ & --- & --- & $>20$~GeV & $>12$~GeV & --- \\  
$m_T(\ell ,\eslt )-veto$ & --- & --- & 60-85~GeV & 60-85~GeV & --- \\  
$OS/SF\ veto$ & no & no & no & no & yes \\
\end{tabular}
\end{center}
\end{table}
\begin{table}
\begin{center}
\caption{Standard Model backgrounds (fb) to the Tevatron SUSY trilepton
signal for the hard as well as for the soft cuts listed in Table~I. }
\bigskip
\begin{tabular}{lccccc}
BG & HC1 & SC1 & SC2 & HC2 & SC3 \\
\hline
$WZ\ (Z\to\tau\bar{\tau})$ & $0.175\pm 0.005$ & $0.40\pm 0.01$  & 
$0.28\pm 0.01$ & $0.16\pm 0.004$ & $0.106\pm 0.004$ \\
$W^*Z^*,W^*\gamma^*\to\ell\ell\bar{\ell}$ & $1.70\pm 0.05$ & $22.0\pm 2.0$  
& $0.21\pm 0.02$ & $0.09\pm 0.01$ & 0 \\
$W^*Z^*,W^*\gamma^*\to\ell\ell'\bar{\ell}'$ & $2.43\pm 0.04$ & $14.6\pm 0.4$  
& $0.48\pm 0.02$ & $0.22\pm 0.003$ & 0 \\
$t\bar{t}$ & $<0.003$ & $0.14\pm 0.006$ & $0.04\pm 0.01$ 
& $0.003\pm 0.003$ & $<0.003$ \\
$Z^*Z^*$       & $0.008\pm 0.001$ & $0.07\pm 0.002$ & $0.04\pm 0.001$ 
& $0.02\pm 0.001$ & $0.010\pm 0.001$ \\
$total$    & $4.31$ & $37.2$ & $1.05$ & 0.49 & 0.116 \\
$5\sigma /5\ ev.  (2\ fb^{-1})$ & $7.3$ & $21.6$ & $3.62$ & 2.5 & 2.5 \\
$5\sigma (25\ fb^{-1})$ & $2.1$ & $6.1$ & $1.02$ & 0.70 & 0.34 \\
$3\sigma (25\ fb^{-1})$ & $1.24$ & $3.7$ & $0.61$ & 0.42 & 0.20  \\
\end{tabular}
\end{center}
\end{table}
\begin{table}
\begin{center}
\caption{Parameter space choices, sparticle masses and
total signal cross sections for the five case studies in Section~IV.
We also list the fractional contribution to the signal from
various subprocesses.
We take $m_t=175$ GeV.}

\bigskip
\begin{tabular}{lccccc}
case & $(1)$ & $(2)$ & $(3)$ & $(4)$ & $(5)$ \\
\hline
$m_0$ & 100 & 140 & 200 & 250 & 150 \\
$m_{1/2}$ & 200 & 175 & 140 & 150 & 300 \\
$A_0$ & 0 & 0 & -500 & -600 & 0 \\
$\tan\beta$ & 3 & 35 & 35 & 3 & 30 \\
$m_{H_u},m_{H_d}$ & -- & -- & -- & -- & 500,500 \\
$m_{\tg}$ & 508 & 455 & 375 & 403 & 734 \\
$m_{\tq}$ & 450 & 410 & 370 & 415 & 650 \\
$m_{\tst_1}$ & 306 & 297 & 153 & 134 & 440 \\
$m_{\tb_1}$  & 418 & 329 & 213 & 346 & 566 \\
$m_{\tw_1}$ & 141 & 126 & 106 & 109 & 100 \\
$m_{\tz_1}$ & 76 & 69 & 56 & 57 & 80 \\
$m_{\tz_2}$ & 143 & 127 & 107 & 111 & 124 \\
$m_{\tz_3}$ & 316 & 252 & 296 & 373 & 141 \\
$m_{\tell_R}$ & 132 & 162 & 212 & 260 & 195 \\
$m_{\tell_L}$ & 180 & 194 & 229 & 275 & 266 \\
$m_{\ttau_1}$ & 131 & 104 & 88 & 257 & 132 \\
$m_{h}$ & 99 & 110 & 112 & 104 & 115 \\
$\mu$ & 312 & 241 & 286 & 369 & -110 \\

$\sigma_{tot.}(fb)    $ & 404 & 653 & 2712 & 3692 & 1393 \\
$\tg ,\tq (\%)        $ & 4.3 & 6.6 & 50.4 & 66.2 & 0.01 \\
$\tg\tx ,\tq\tx (\%)  $ & 2.4 & 3.6 & 2.9  & 1.2  & 0.01 \\
$\tx\tx (\%)          $ & 85.0& 85  & 45.7 & 32.6 & 99.5 \\
$\tell\tell (\%)      $ & 8.3 & 4.7 & 1.0  & 0.04 & 0.4  \\
$\tst_1\tst_1 (\%)    $ & 1.8 & 1.5 & 41   & 65   & 0.01 \\
$\tw_1\tz_2 (\%)      $ & 43.8& 45  & 26.5 & 18   & 16.7 \\
$\tw_1\twb_1 (\%)     $ & 33.5& 33  & 17.6 & 13   & 24.6 \\
\end{tabular}
\end{center}
\end{table}
\begin{table}
\begin{center}
\caption{SUSY $3\ell$ signal (fb) for hard and soft cuts at the
Tevatron for Cases 1--5 described in Section~IV and for Cases A, B, and
C corresponding to Figs.~\ref{PTL1}--\ref{PTL3}. The two columns for SC1
cuts refer to cross sections with and without the inclusion of decay
matrix elements as discussed in the text.}
\bigskip
\begin{tabular}{lcccccc}
case & HC1 & SC1 (no ME) & SC1 (ME included) & SC2 & HC2 & SC3 \\
\hline
(1) & $3.3\pm 0.2$ & $13.1\pm 0.5$ & $13.1\pm 0.5$ & $6.9\pm 0.3$ 
& $3.5\pm 0.2$ & $0.11\pm 0.04$ \\
(2) & $0.17\pm 0.04$ & $1.6\pm 0.1$ & $1.6\pm 0.1$ & $0.8\pm 0.1$ 
& $0.26\pm 0.05$ & $0.26 \pm 0.06$ \\
(3) & $0.30\pm 0.10$ & $2.5\pm 0.3$ & $2.5\pm 0.3$ & $1.2\pm 0.2$ 
& $0.41\pm 0.1$ & $0.6\pm 0.1$ \\
(4) & $1.4\pm 0.3$   & $3.6\pm 0.5$ & $3.3\pm 0.5$ & $2.1\pm 0.4$ 
& $1.2\pm 0.3$ & $0.07\pm 0.07$ \\
(5) & $0.3\pm 0.1$   & $1.4\pm 0.2$ & $1.2\pm 0.2$ & $0.6\pm 0.1$ 
& $0.33\pm 0.1$ & $0.14\pm 0.06$ \\
\hline
(A) & $7.4\pm 0.4$ & $20.6 \pm 0.1$ & $19.6 \pm 0.6$ & $11.4\pm 0.5 $ 
& $8.0\pm 0.4$ & $0.26\pm 0.08$ \\
(B) & $0.16 \pm 0.05$ & 0.55$\pm$ 0.1 & $0.6 \pm 0.1$ & $0.33\pm 0.01$ 
& $0.19\pm 0.05$ & $0.015\pm 0.015$ \\
(C) & $0.16 \pm 0.02$ & 0.44$\pm$0.04 & $0.25 \pm 0.02$ & $0.12\pm 0.02$
& $0.10\pm 0.02$ & $0.01\pm 0.005$ 
\end{tabular}
\end{center}
\end{table}
   
%
%
%

\newpage
\iftightenlines\else\newpage\fi
\iftightenlines\global\firstfigfalse\fi
\def\dofig#1#2{\epsfxsize=#1\centerline{\epsfbox{#2}}}

\begin{figure}
\dofig{5in}{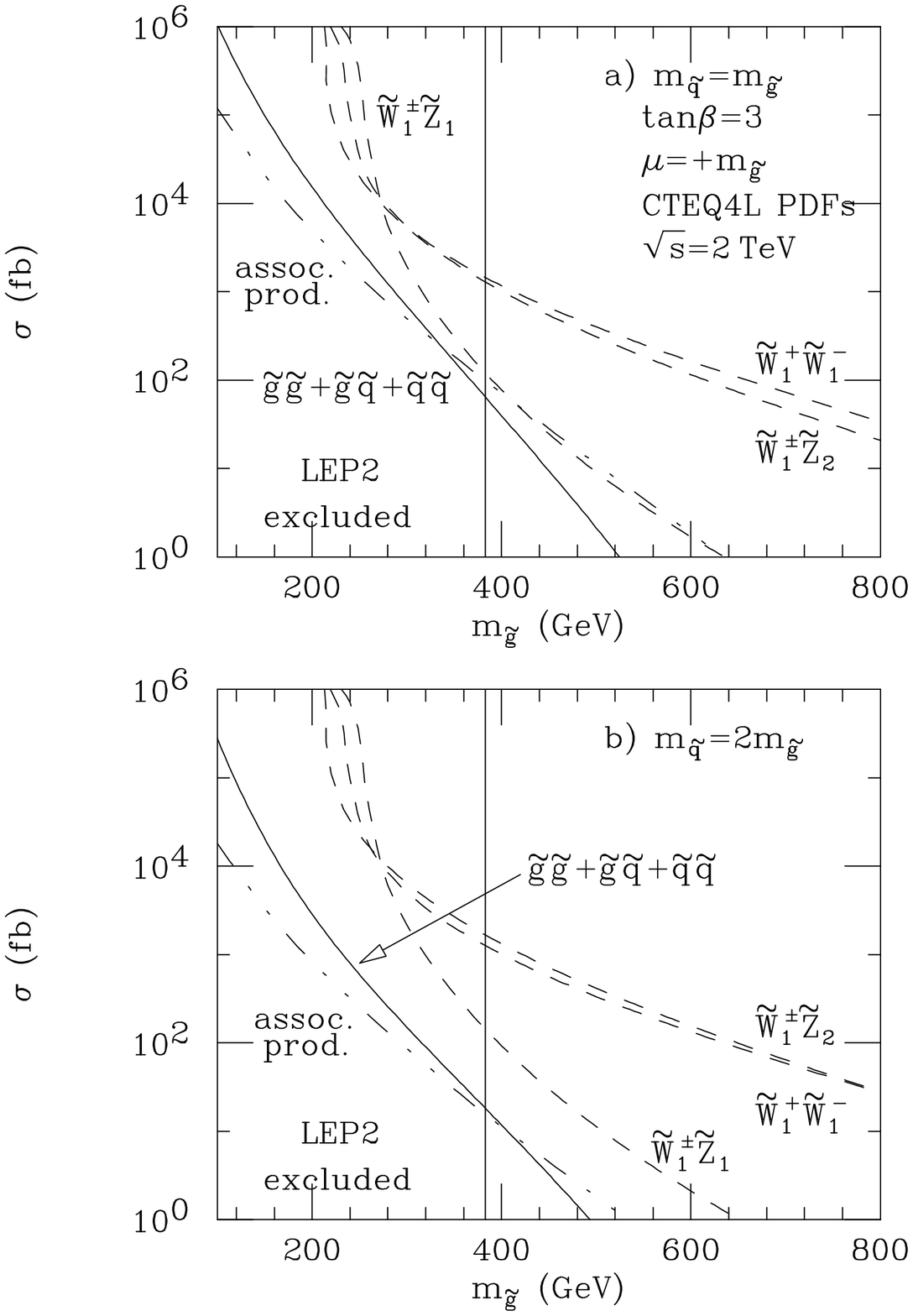}
\caption[]{
Sparticle production cross sections as a function of $m_{\tilde g}$
for $\mu=+m_{\tilde g}$ and $\tan\beta =3$.}
\label{SIGA}
\end{figure}
%
\begin{figure}
\dofig{5in}{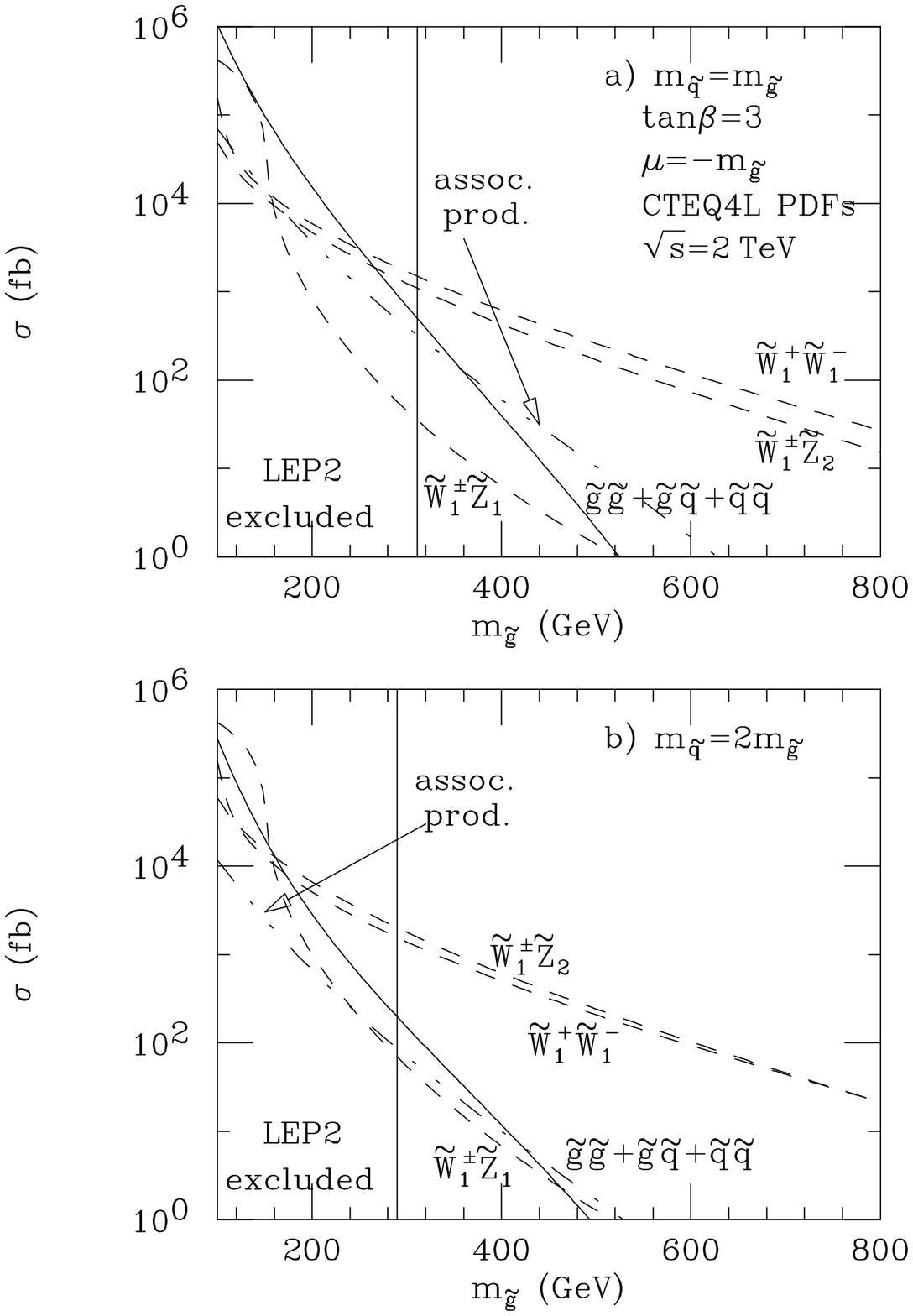}
\caption[]{
Sparticle production cross sections as a function of $m_{\tilde g}$
for $\mu=-m_{\tilde g}$ and $\tan\beta =3$.}
\label{SIGB}
\end{figure}
%
\begin{figure}
\dofig{5in}{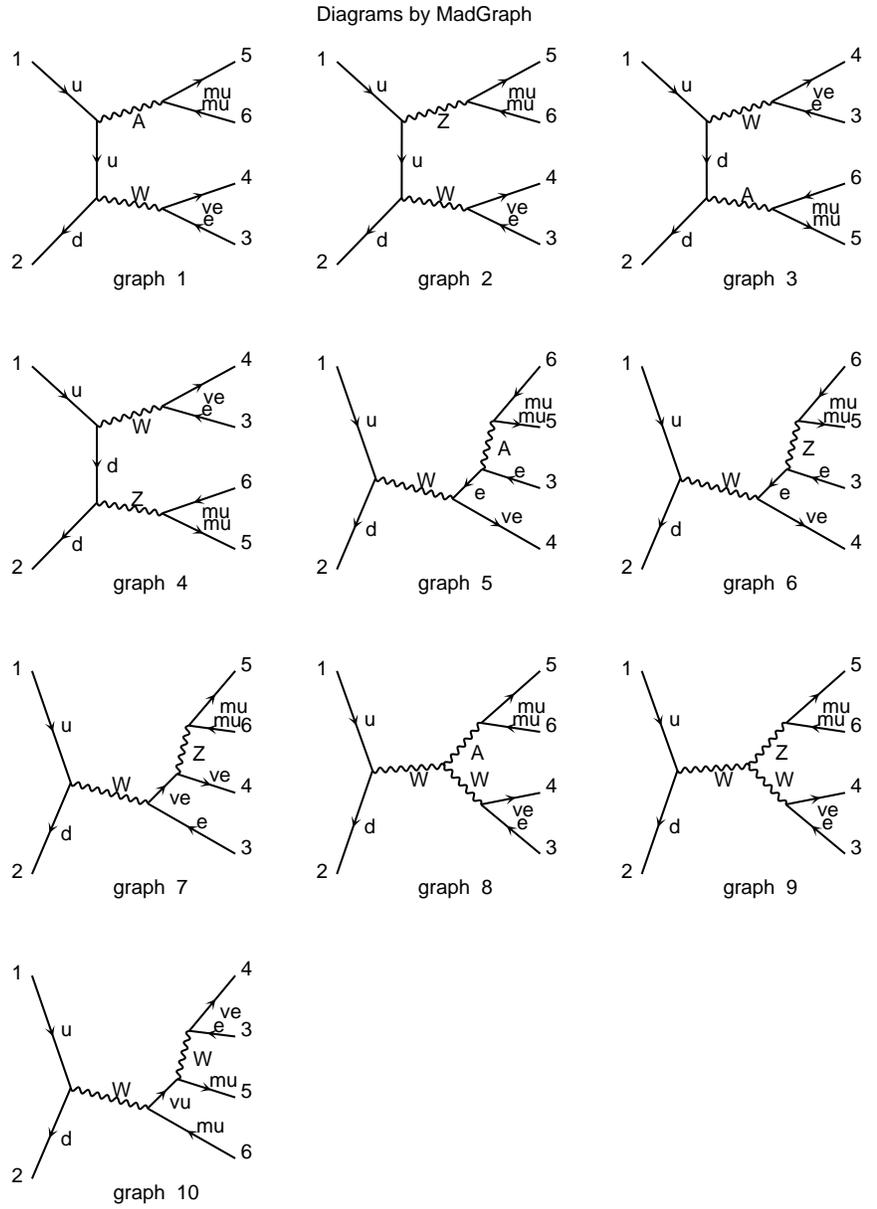}
\caption[]{
Feynman diagrams contributing to $W^*\gamma^*$, $W^*Z^*$ background.}
\label{DIAGRAMS}
\end{figure}
%
\begin{figure}
\dofig{4in}{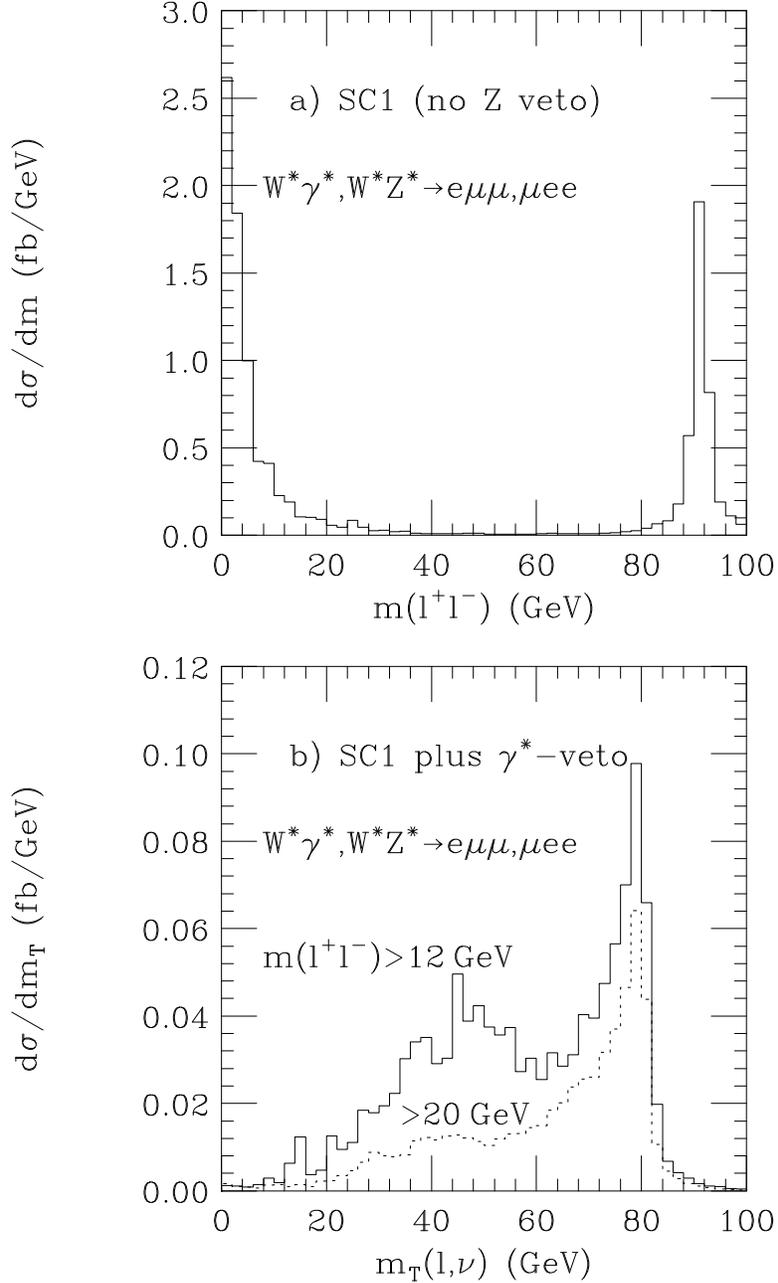}
\caption[]{
{\it a}) Distribution of same-flavor, opposite sign dilepton mass from
$W^*\gamma^*$, $W^*Z^*\to e\mu\mu ,\ \mu ee$ background after
cuts SC1, but with the $Z$-mass veto removed.
In {\it b}), we show the distribution in transverse mass from the 
same background with cuts SC1, including the $Z$ and $\gamma$ veto.}
\label{MLLWZ}
\end{figure}
%
\begin{figure}
\dofig{4in}{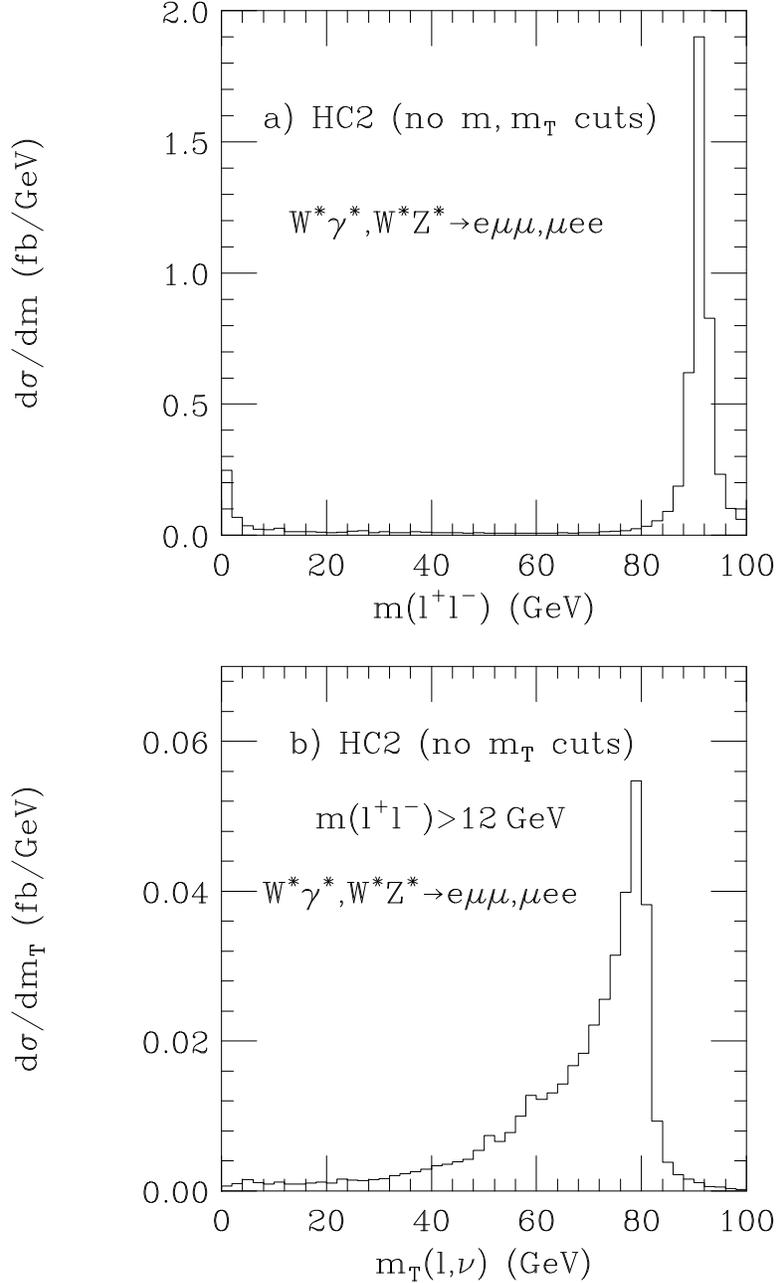}
\caption[]{
{\it a}) Distribution of same-flavor, opposite sign dilepton mass from
$W^*\gamma^*$, $W^*Z^*\to e\mu\mu ,\ \mu ee$ background after
cuts HC2, but with the $m(\ell^+\ell^- )$ and $m_T$ cuts removed.
In {\it b}), we show the distribution in transverse mass from the 
same background with cuts HC2, including the $Z$ and $\gamma$ veto, but
without the $m_T$ cut.}
\label{MLLWZH}
\end{figure}
%
\begin{figure}
\dofig{3in}{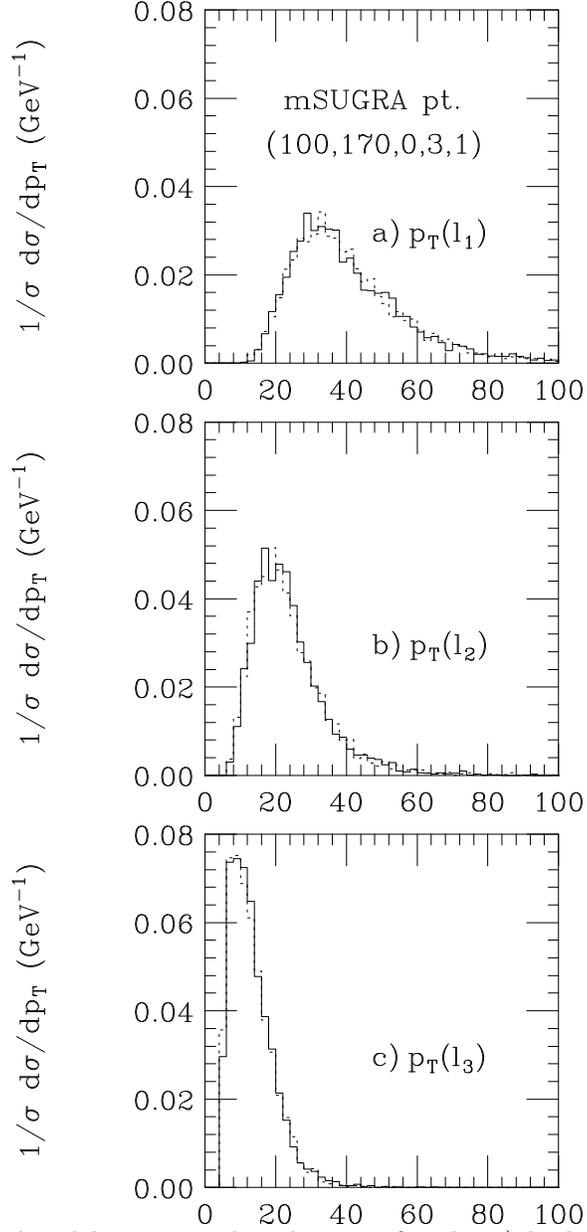}
\caption[]{Normalized isolated lepton $p_T$ distributions for the 
{\it a}) highest, {\it b}) second highest and {\it c}) lowest $p_T$
lepton in trilepton events for the listed mSUGRA point (case A)
after inclusive
soft cuts SC1 listed in Table~I. 
The dashed histogram denotes the case of phase space decays, while the solid
histogram denotes the case with exact three-body decay matrix elements.}
\label{PTL1}
\end{figure}
%
\begin{figure}
\dofig{3in}{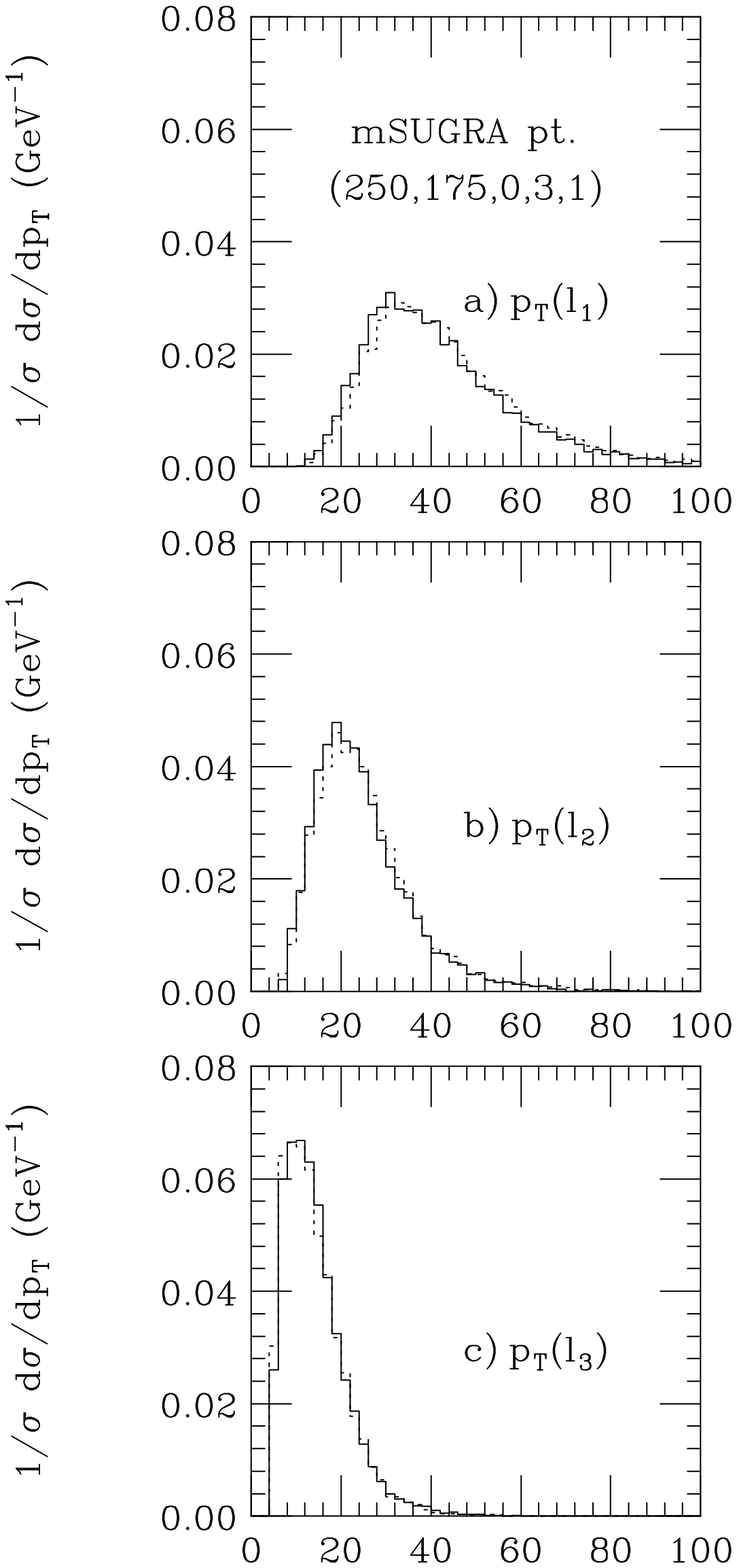}
\caption[]{Same as Fig.~\ref{PTL1}, except with mSUGRA 
parameters for case B.}
\label{PTL2}
\end{figure}
%
\begin{figure}
\dofig{3in}{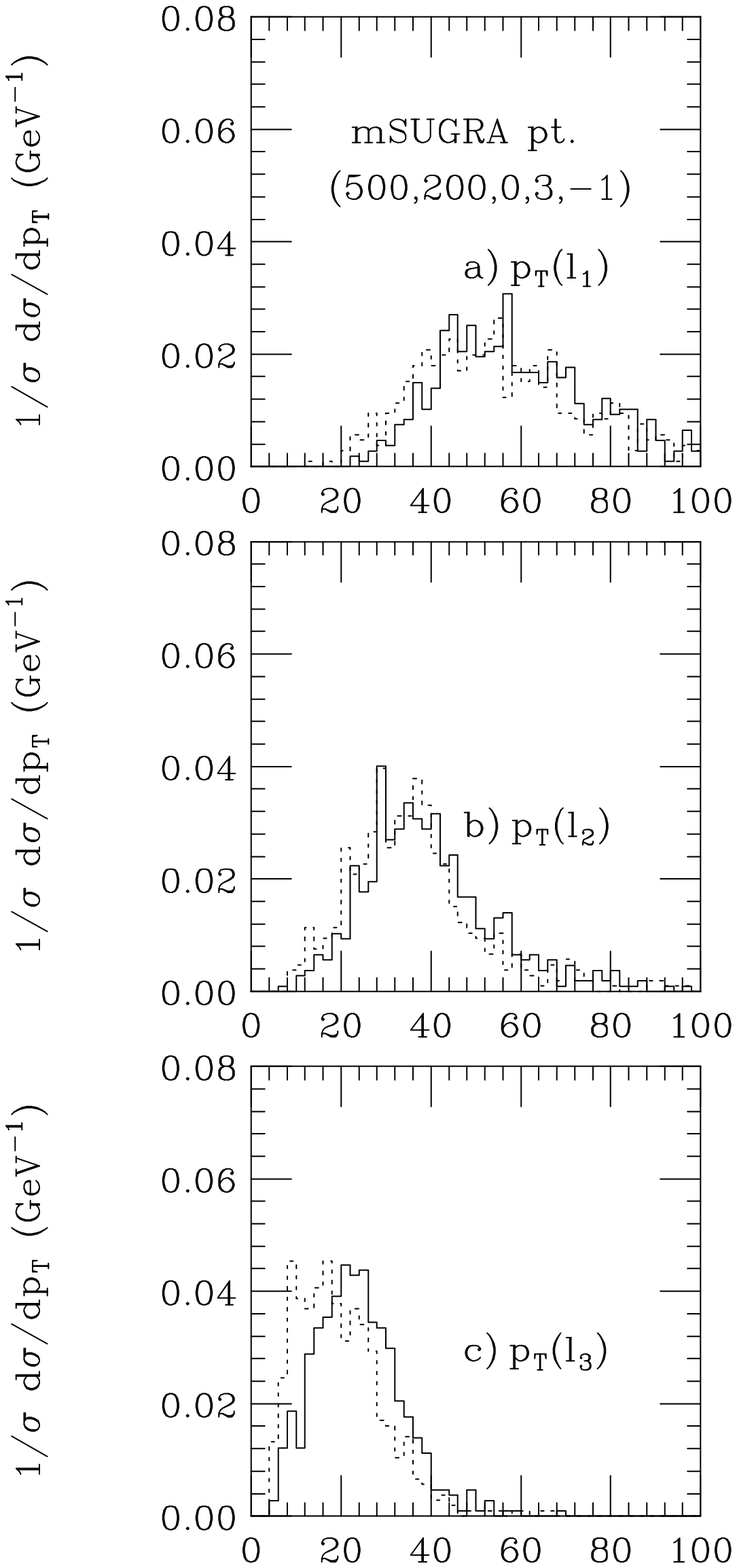}
\caption[]{Same as Fig.~\ref{PTL1}, except with mSUGRA 
parameters for case C.}
\label{PTL3}
\end{figure}
%
\begin{figure}
\dofig{5in}{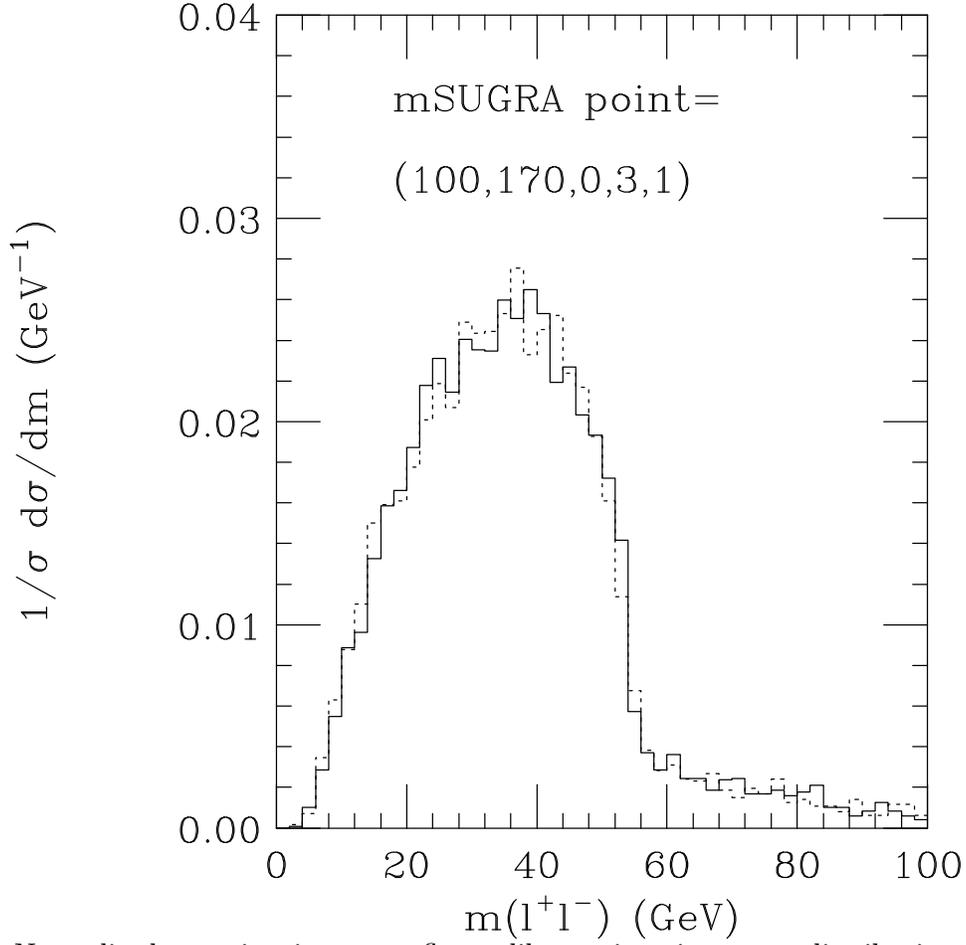}
\caption[]{Normalized opposite sign, same-flavor dilepton invariant mass 
distribution after inclusive soft cuts SC1 (other than the $Z$-veto) for
the listed mSUGRA point, case A. We show the result using only 
phase space for the decay matrix element (dashes), and 
the result using the exact decay squared matrix element (solid).}
\label{POINT1}
\end{figure}
%
\begin{figure}
\dofig{5in}{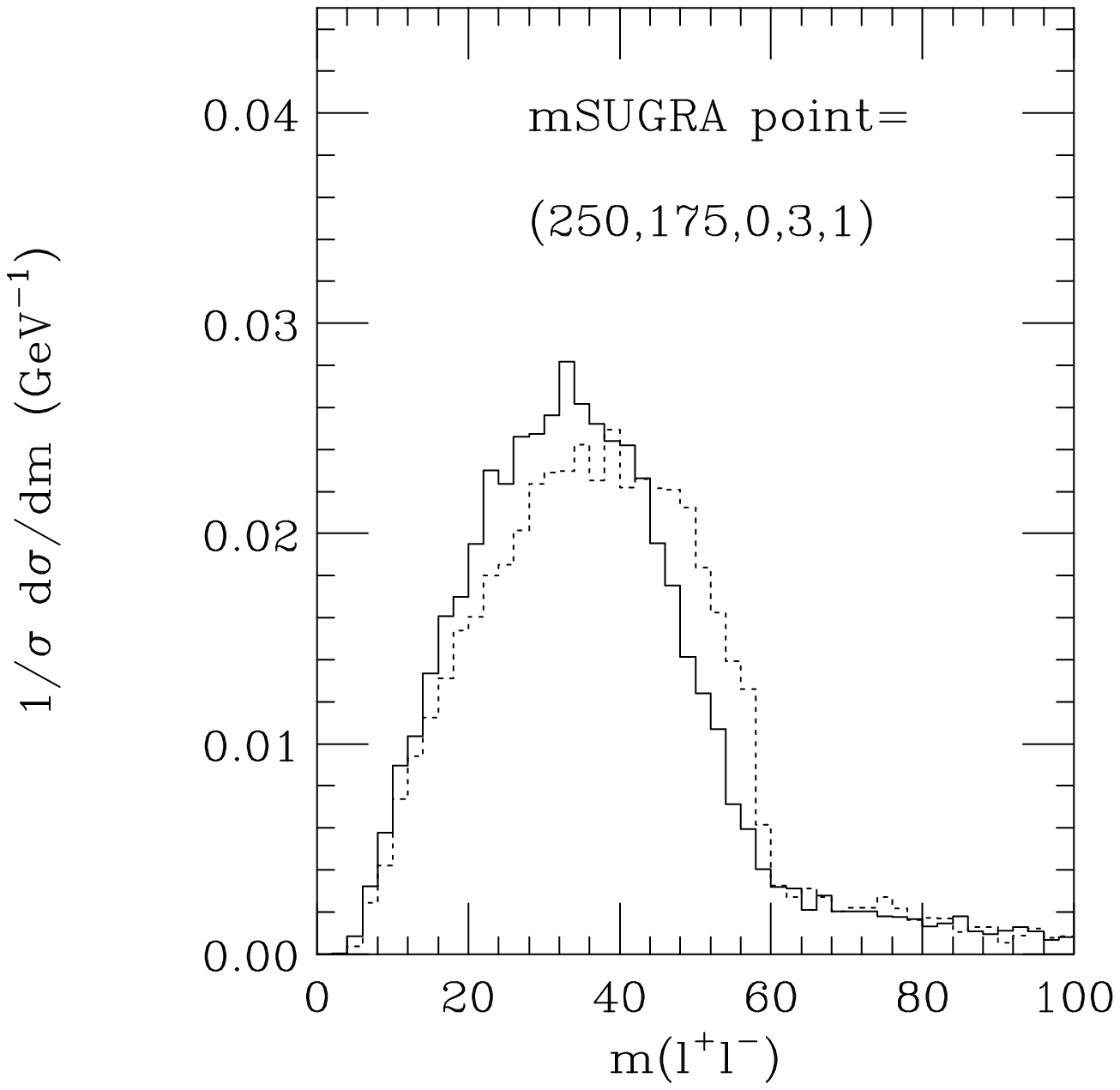}
\caption[]{The same as Fig.~\ref{POINT1} except for case B where the model
parameters are as listed.}
\label{POINT2}
\end{figure}
%
\begin{figure}
\dofig{5in}{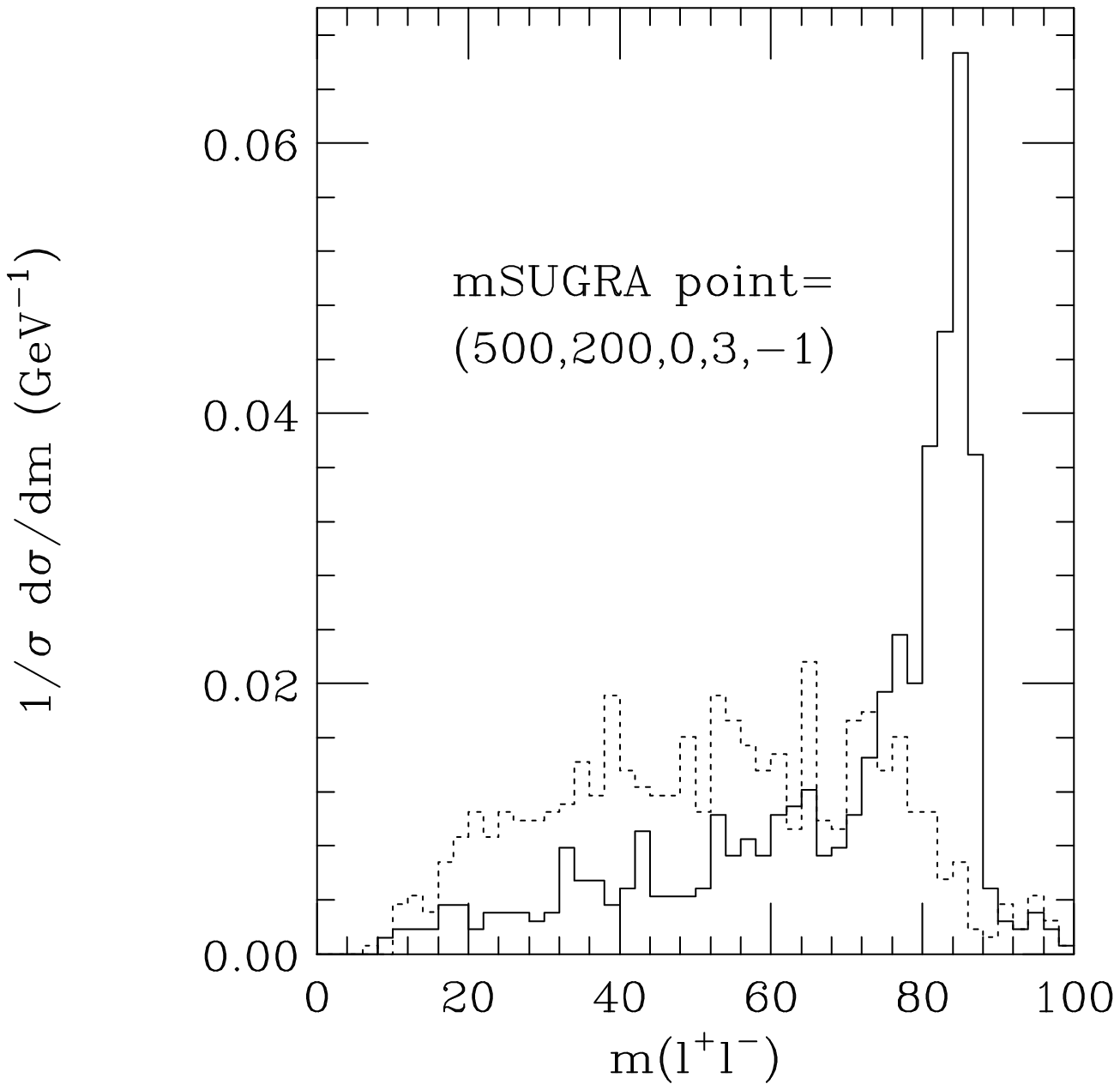}
\caption[]{The same as Fig.~\ref{POINT1} except for case C where the
model parameters are as listed.}
\label{POINT3}
\end{figure}
%
\begin{figure}
\dofig{5in}{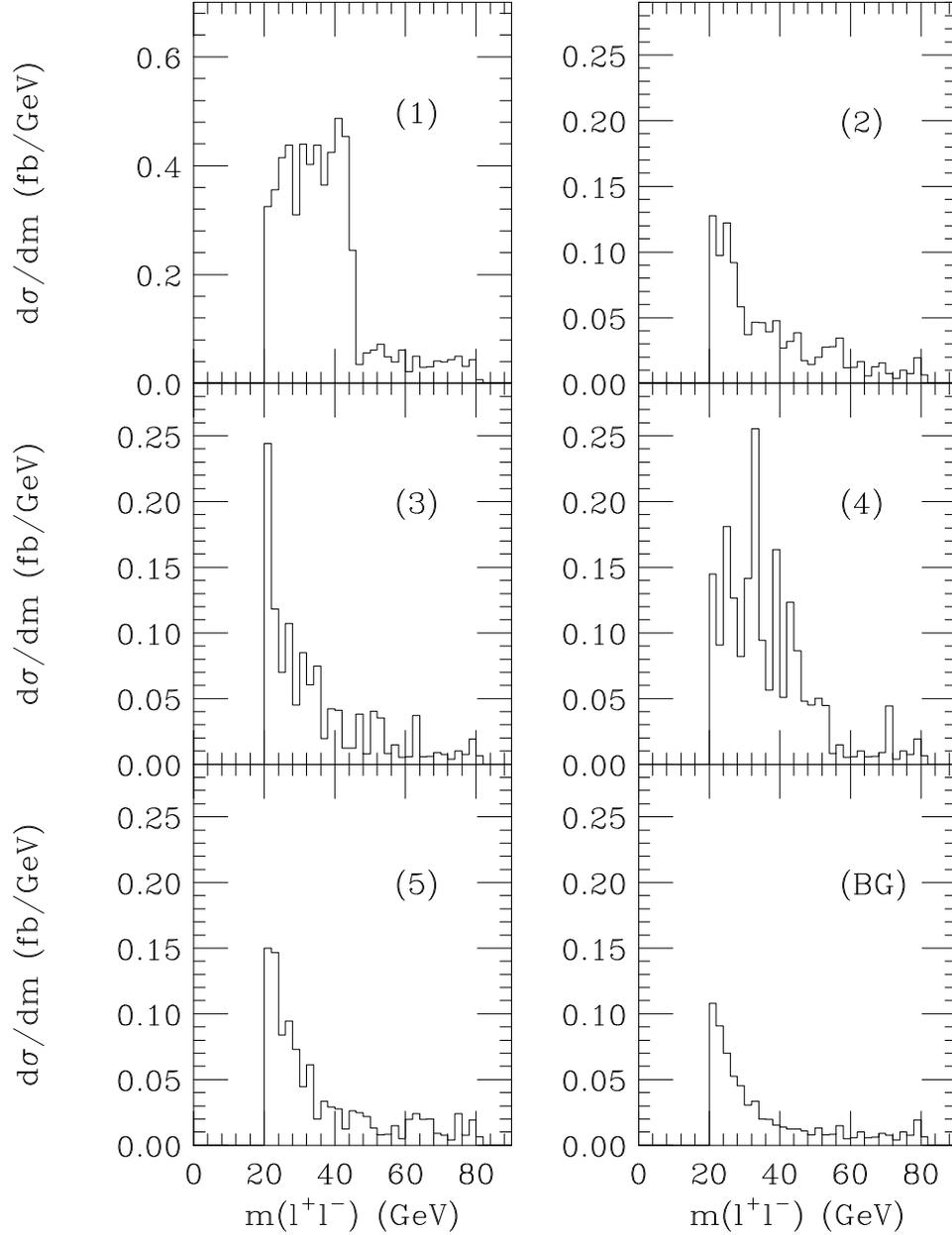}
\caption[]{Opposite sign same-flavor dilepton invariant mass
distributions after cuts SC2 listed in Table~I for the
five case study points introduced in Section~IV of the text.  Each frame
is labelled by the particular case number. The last frame labelled BG
shows the sum of SM backgrounds from Table~II. The
background is included in the histogram for each case study.  The cross
section in each plot is greater than the value listed in Table 4 since
some events can have more than one plot entry.}
\label{MLL}
\end{figure}
%
\begin{figure}
\dofig{6in}{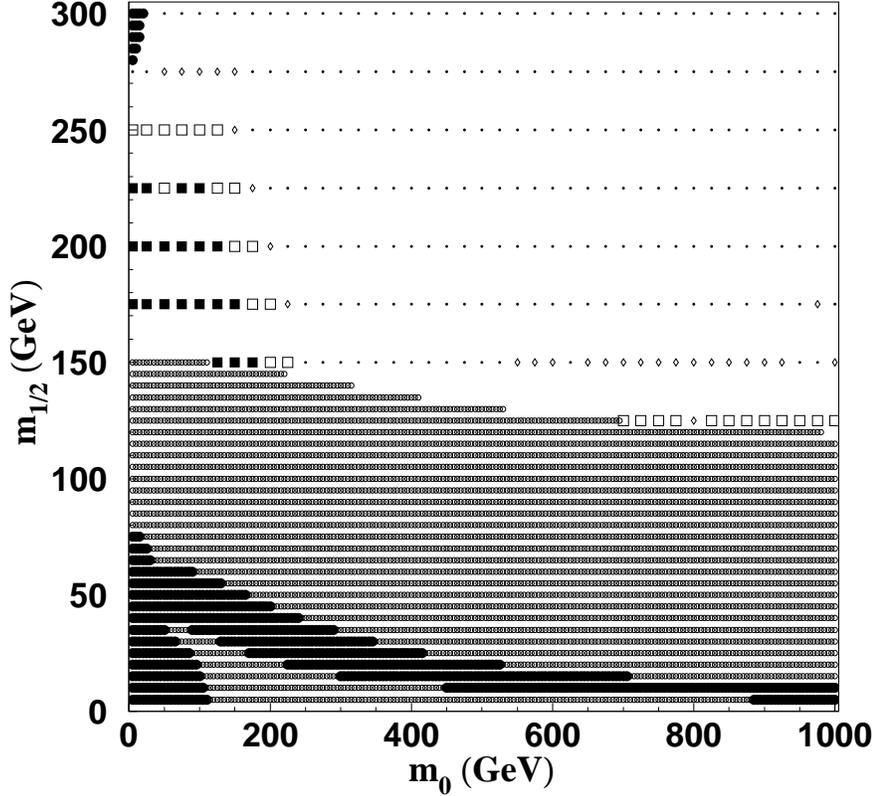}
\caption[]{The reach of the Fermilab Tevatron collider for mSUGRA model
in the $m_0\ vs.\ m_{1/2}$ plane for $A_0=0$, $\tan\beta =3$ and $\mu >0$,
using cuts SC2.
The black shaded regions are theoretically excluded, while the gray
areas are experimentally excluded by sparticle and Higgs boson searches
at LEP2. The black squares denote points accessible to Tevatron 
experiments at the
$5\sigma$ level with just 2 fb$^{-1}$ of data, while open squares
are accessible with 25 fb$^{-1}$. Points denoted by diamonds are accessible 
at the $3\sigma$ level with 
25 fb$^{-1}$ of integrated luminosity.}
\label{PAM3N}
\end{figure}
%
\begin{figure}
\dofig{6in}{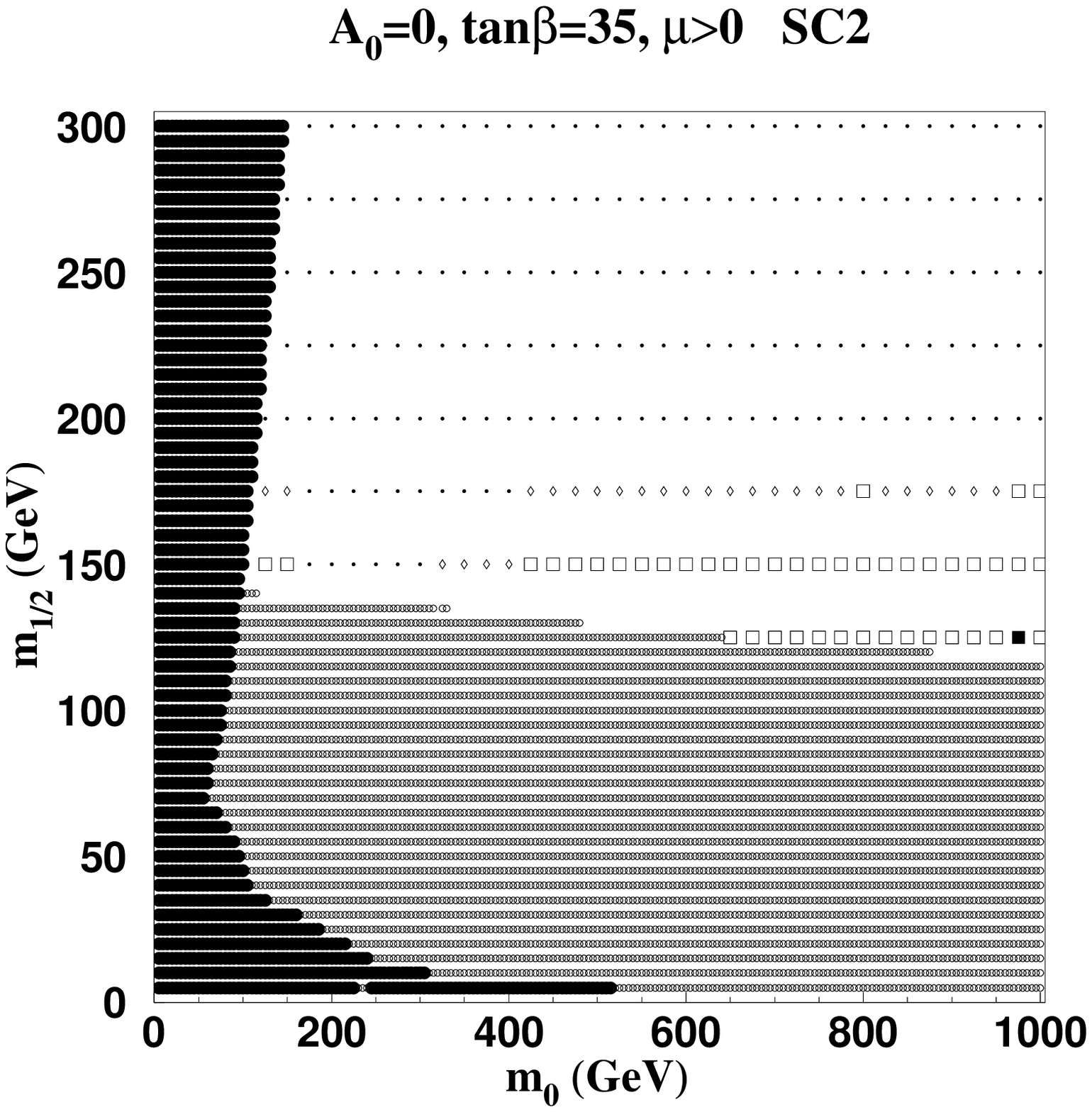}
\caption[]{Same as Fig.~\ref{PAM3N}, except for $\tan\beta =35$.}
\label{PAM35N}
\end{figure}
%
\begin{figure}
\dofig{6in}{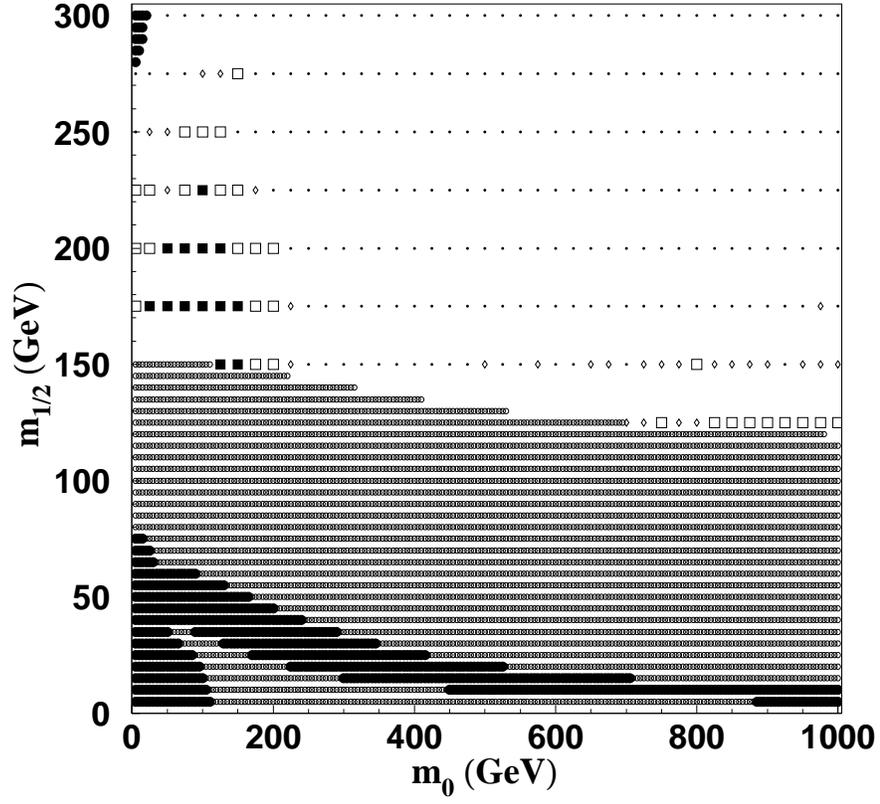}
\caption[]{Same as Fig.~\ref{PAM3N}, except for cuts HC2.}
\label{PAM3NH}
\end{figure}
%
\begin{figure}
\dofig{6in}{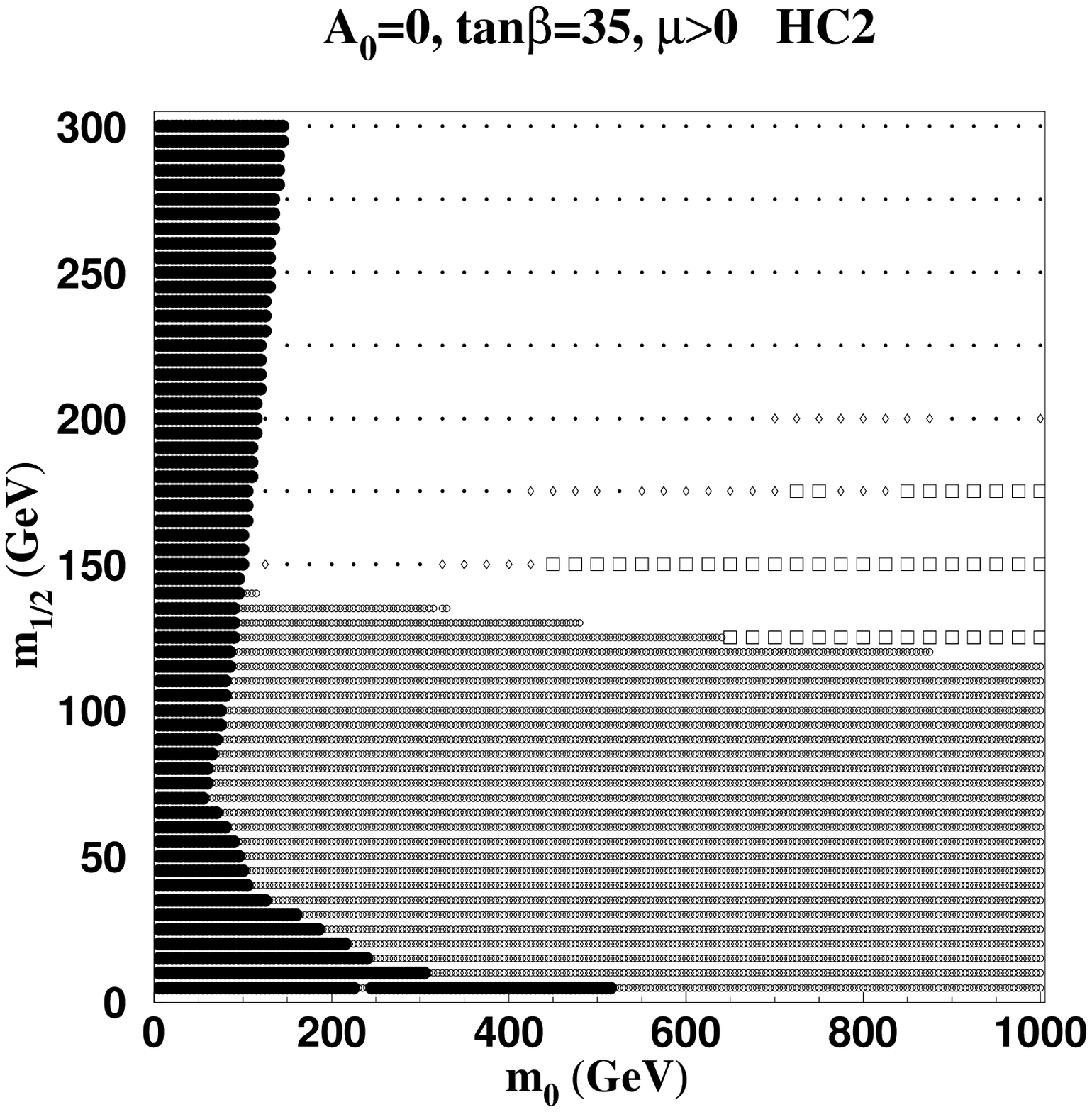}
\caption[]{Same as Fig.~\ref{PAM3NH}, except for $\tan\beta =35$.}
\label{PAM35H}
\end{figure}
\end{document}